\documentclass[aps,prb,floatfix,twocolumn,showpacs]{revtex4}%
\usepackage{amsmath}
\usepackage{graphicx}
\usepackage{amsfonts}
\usepackage{amssymb}%
\providecommand{\U}[1]{\protect\rule{.1in}{.1in}}

\begin{document}
\title{Spin-dependent boundary conditions for isotropic superconducting Green's functions}
\author{Audrey Cottet$^{1,2}$, Daniel Huertas-Hernando$^{3}$, Wolfgang Belzig$^{4}$
and Yuli V. Nazarov$^{5}$}
\affiliation{{$^{1}$ Ecole Normale Sup\'{e}rieure, Laboratoire Pierre Aigrain, 24 rue
Lhomond, F-75231 Paris Cedex 05, France}}
\affiliation{{$^{2}$ CNRS UMR8551, Laboratoire associ\'{e} aux universit\'{e}s Pierre et
Marie Curie et Denis Diderot, France}}
\affiliation{{$^{3}$ Department of Physics, Norwegian University of Science and Technology,
N-7491 Trondheim, Norway}}
\affiliation{{$^{4}$ Department of Physics, University of Konstanz, D-78457 Konstanz,
Germany}}
\affiliation{{$^{5}$ Kavli Institute of NanoScience, Delft University of Technology,
NL-2628 CJ Delft, The Netherlands}}
\date{\today}


\begin{abstract}
The quasiclassical theory of superconductivity provides the most successful
description of diffusive heterostructures comprising superconducting elements,
namely, the Usadel equations for isotropic Green's functions. Since the
quasiclassical and isotropic approximations break down close to interfaces,
the Usadel equations have to be supplemented with boundary conditions for
isotropic Green's functions (BCIGF), which are not derivable within the
quasiclassical description. For a long time, the BCIGF were available only for
spin-degenerate tunnel contacts, which posed a serious limitation on the
applicability of the Usadel description to modern structures containing
ferromagnetic elements. In this article, we close this gap and derive
spin-dependent BCIGF for a contact encompassing superconducting and
ferromagnetic correlations. This finally justifies several simplified versions
of the spin-dependent BCIGF, which have been used in the literature so far. In
the general case, our BCIGF are valid as soon as the quasiclassical isotropic
approximation can be performed. However, their use require the knowledge of
the full scattering matrix of the contact, an information usually not
available for realistic interfaces. In the case of a weakly polarized tunnel
interface, the BCIGF can be expressed in terms of a few parameters, i.e. the
tunnel conductance of the interface and five conductance-like parameters
accounting for the spin-dependence of the interface scattering amplitudes. In
the case of a contact with a ferromagnetic insulator, it is possible to find
explicit BCIGF also for stronger polarizations. The BCIGF derived in this
article are sufficienly general to describe a variety of physical situations
and may serve as a basis for modelling realistic nanostructures.

\end{abstract}
\maketitle


\section{Introduction\label{introduction}}

The quantum mechanical spin degree of freedom is widely exploited to control
current transport in electronic circuits nowadays. For instance, the readout
of magnetic hard disks is based on the giant magnetoresistance effect, which
provides the possibility to tune the conductance of e.g.~a ferromagnet/normal
metal/ferromagnet ($F/N/F$) trilayer by changing the magnetizations of the two
$F$ layers from a parallel to an antiparallel configuration\cite{GMR}.
However, many functionalities of hybrid circuits enclosing ferromagnetic
elements remain to be explored. Presently, non-collinear spin transport is
triggering an intense activity, due to spin-current induced magnetization
torques\cite{Slonczewski}, which offer new possibilities to build non-volatile
memories\cite{RalphStiles}. Another interesting possibility is to include
superconducting elements in hybrid circuits. When a $N$ layer is connected to
a BCS superconductor ($S$), the singlet electronic correlations characteristic
of $S$ can propagate into $N$ because electrons and holes with opposite spins
are coupled coherently by Andreev reflections occurring at the $S/N$
interface\cite{RefAndreev}. This so-called "superconducting proximity effect"
is among other responsible for strong modifications of the density of states
of $N$\cite{RefMcMillan}. In a ferromagnet ($F$), the ferromagnetic exchange
field $E_{ex}$, which breaks the symmetry between the two spin bands, is
antagonistic to the Bardeen-Cooper-Schrieffer(BCS)-type singlet
superconducting order. However, this does not exclude the superconducting
proximity effect. First, when the magnetization direction is uniform in a
whole $S/F$ circuit, superconducting correlations can occur between electrons
and holes from opposite spin bands, like in the $S/N$ limit. These
correlations propagate on a characteristic distance limited by the
ferromagnetic coherence length $\sqrt{\hbar D/E_{ex}}$, where $D$ is the
diffusion coefficient. Furthermore, $E_{ex}$ produces an energy shift between
the correlated electrons and holes in the opposite spin bands, which leads to
spatial oscillations of the superconducting order parameter in $F$%
\cite{Buzdin1982}, as recently observed\cite{TakisN,Ryazanov,TakisI}. These
oscillations allow to build new types of electronic devices, such as Josephson
junctions with negative critical currents\cite{Guichard}, which promise
applications in the field of superconducting circuits\cite{Ioffe,Taro}.
Secondly, when the circuit encloses several ferromagnetic elements with
noncollinear magnetizations, spin precession effects allow the existence of
superconducting correlations between equal spins\cite{Bergeret}. These
correlations are expected to propagate in a $F$ on a distance much longer than
opposite-spin correlations. This property could be used e.g. to engineer a
magnetically switchable Josephson junction. These and many more effects have
been reviewed recently \cite{Golubov,Buzdin}.

To model the behavior of superconducting hybrid circuits, a proper description
of the interfaces between the different materials is crucial. This article
focuses on the so-called diffusive limit, which is appropriate for most
nanostructures available nowadays. In this limit, a nanostructure can be
separated into interfaces (or contacts) and regions characterized by isotropic
Green functions $\check{G}$, which do not depend on the direction of the
momentum but conserve a possible dependence on spatial coordinates. The
spatial evolution of the isotropic Green functions $\check{G}$ is described by
Usadel equations \cite{Usadel}. One needs boundary conditions to relate the
values of $\check{G}$ at both sides of an interface. For a long time, the only
boundary conditions for isotropic Green's functions (BCIGF) available were
spin-independent BCIGF derived for a $S/N$ tunnel contact\cite{Kuprianov}. The
only interfacial parameter involved in these BCIGF was the tunnel conductance
$G_{T}$ of the contact. Such a description is incomplete for a general
diffusive spin-dependent interface. Spin-dependent $S/F$ boundary conditions
have been first introduced in the ballistic
regime\cite{Millis,Tokuyasu,Kopu,Zhao}. Recently, many references have used
spin-dependent
BCIGF\cite{Dani1,Dani2,Cottet1,Cottet2,Cottet3,Linder1,Morten1,Morten2,Braude,Linder2,DiLorenzo}
to study the behavior of hybrid circuits enclosing BCS superconductors,
ferromagnetic insulators, ferromagnets, and normal metals. These BCIGF, which
have been first introduced in Ref.~\onlinecite{Dani1}, include the $G_{T}$
term of Ref.~\onlinecite{Kuprianov}. They furthermore take into account the
spin-polarization of the interface tunnel probabilities through a $G_{MR}$
term, and the spin-dependence of interfacial scattering phase shifts through
$G_{\phi}$ terms. It has been shown that the $G_{MR}$ and $G_{\phi}$ terms
lead to a rich variety of effects. First, the $G_{\phi}$ terms can produce
effective Zeeman fields inside thin superconducting or normal metal
layers\cite{Dani1,Dani2,Cottet2}, an effect which could be used e.g. to
implement an absolute spin-valve\cite{Dani1}. In thick superconducting layers,
this effect is replaced by spin-dependent resonances occurring at the edges of
the layers\cite{Cottet3}. Secondly, the $G_{\phi}$ terms can shift the spatial
oscillations of the superconducting order parameter in
ferromagnets\cite{Cottet2,Cottet1,Cottet3}. Thirdly, the $G_{\phi}$ term can
produce superconducting correlations between equal spins, e.g. in a circuit
enclosing a BCS superconductor and several ferromagnetic insulators magnetized
in noncollinear directions\cite{Braude}. The $G_{MR}$ terms have been taken
into account for a chaotic cavity connected to a superconductor and several
ferromagnets\cite{Morten1,Morten2}. In this system, crossed Andreev
reflections and direct electron transfers are responsible for nonlocal
transport properties. The ratio between these two kinds of processes, which
determines e.g. the sign of the nonlocal conductance\cite{Falci01,Sanchez03},
can be controlled through the relative orientation of the ferromagnets magnetizations.

In this article, we present a detailed derivation of the spin-dependent BCIGF
based on a scattering description of interfaces. Our results thus provide a
microscopic basis for all future investigations of ferromagnet-superconductor
diffusive heterostructures taking into account the spin-dependent interface
scattering. To make the BCIGF comprehensive and of practical value, we make a
series of sequential assumptions, starting from very general to more and more
restrictive hypotheses. In a first part, we assume that the contact is fully
metallic, i.e. it connects two conductors which can be superconductors,
ferromagnets or normal metals. We consider ferromagnets with exchange fields
much smaller than their Fermi energies, as required for the applicability of
the quasiclassical isotropic description. We assume that the contact
nevertheless produces a spin-dependent scattering due to a spin-dependent
interfacial barrier $\bar{V}_{b}$. In this case, we establish general BCIGF
which require the knowledge of the full contact scattering matrix. Then, we
assume that the contact locally conserves the transverse channel index
(specular hypothesis) and spins collinear to the contact magnetization. In the
tunnel limit, assuming $\bar{V}_{b}$ is weakly spin-dependent, we find that
the BCIGF involve the $G_{T}$, $G_{MR}$, and $G_{\phi}$ terms used in
Refs.~\onlinecite{Cottet1,Cottet2,Cottet3,Linder1,Morten1,Morten2,Braude,Linder2,DiLorenzo},
plus additional $G_{\chi}$ terms which are usually disregarded. In a second
part, we study a specular contact connecting a metal to a ferromagnetic
insulator ($FI$). If we assume a weakly spin-dependent interface scattering,
we obtain the BCIGF used in Refs.~\onlinecite{Dani1,Dani2}. We also present
BCIGF valid beyond this approximation. Note that the various BCIGF presented
in this article can be applied to noncollinear geometries.

Most of the literature on superconducting hybrid circuits uses a spatially
continuous description, i.e., in each conductor, the spatial dependence of the
Green's function $\check{G}$ is explicitly taken into account. The BCIGF
presented in this article can also be used in the alternative approach of the
so-called circuit theory. This approach is a systematic method to describe
multiterminal hybrid structures, in order to calculate average transport
properties\cite{RefCircuit,Braatas,RefYuli} but also current
statistics\cite{yuli:99,wn01}. It relies on the mapping of a real geometry
onto a topologically equivalent circuit represented by finite elements. The
circuit is split up into reservoirs (voltage sources), connectors (contacts,
interfaces) and nodes (small electrodes) in analogy to classical electric
circuits. Each reservoir or node is characterized by an isotropic Green's
function $\check{G}$ without spatial dependence, which plays the role of a
generalized potential. One can define matrix currents, which contain
information on the flows of charge, spin, and electron/hole coherence in the
circuit. Circuit theory requires that the sum of all matrix currents flowing
from the connectors into a node is balanced by a \textquotedblleft
leakage\textquotedblright\ current which accounts for the non-conservation of
electron/hole coherence and spin currents in the node. This can be seen as a
generalized Kirchhoff's rule and completely determines all the properties of
the circuit. So far, circuit theory has been developed separately for
$F/N$\cite{RefCircuit} and $S/N$ circuits\cite{RefYuli}. Throughout this
article, we express the BCIGF in terms of matrix currents. Our work thus
allows a straightforward generalization of circuit theory to the case of
multiterminal circuits which enclose superconductors, normal metals,
ferromagnets and ferromagnetic insulators, in a possibly noncollinear geometry.

This article is organized as follows. We first consider the case of a metallic
contact, i.e. a contact between two conductors. Section~\ref{general} defines
the general and isotropic Green's functions $\mathbb{G}$ and $\check{G}$ used
in the standard description of hybrid circuits encompassing BCS
superconductors. Section \ref{ballistic2} introduces the ballistic Green's
function $\tilde{g}$, which we use in our derivation. Section~\ref{ballistic}
discusses the scattering description of the contact with a transfer matrix
$\bar{M}$. Although we consider the diffusive limit, the scattering
description is relevant for distances to the contact shorter than the elastic
mean free path. On this scale, one can use $\bar{M}$\ to relate the left and
right ballistic Green's functions $\tilde{g}_{L}$ and $\tilde{g}_{R}$.
Section~\ref{isotropization} presents an isotropization scheme which accounts
for impurity scattering and leads to the isotropic Green's functions
$\check{G}_{L(R)}$ away from the contact. Section~\ref{generalBC} establishes
the general metallic BCIGF which relate $\check{G}_{L}$, $\check{G}_{R}$ and
$\bar{M}$. Section \ref{WeakFerro} gives more transparent expressions of these
BCIGF in various limits. Section~\ref{SFI} addresses the case of a contact
with a $FI$ side, in analogy with the treatment realized in the metallic case.
Section~\ref{conclusion} concludes. Appendix \ref{Scattering} discusses the
structure of the transfer matrix $\bar{M}$ and Appendix \ref{MatrixCurrent}
gives details on the calculation of the matrix current. Appendix
\ref{NormalState} relates our BCIGF to the equations previously obtained in
the normal-state limit\cite{RefCircuit,Braatas}. Appendix \ref{GorkovToUsadel}
discusses the BCIGF obeyed by the retarded parts of $\check{G}_{L(R)}$ in the
collinear case. For completeness, Appendix \ref{UsadelAppendix} presents the
Usadel equations in our conventions.

\section{General and isotropic Green's functions \label{general}}

\begin{figure}[ptbh]
\includegraphics[width=1\linewidth]{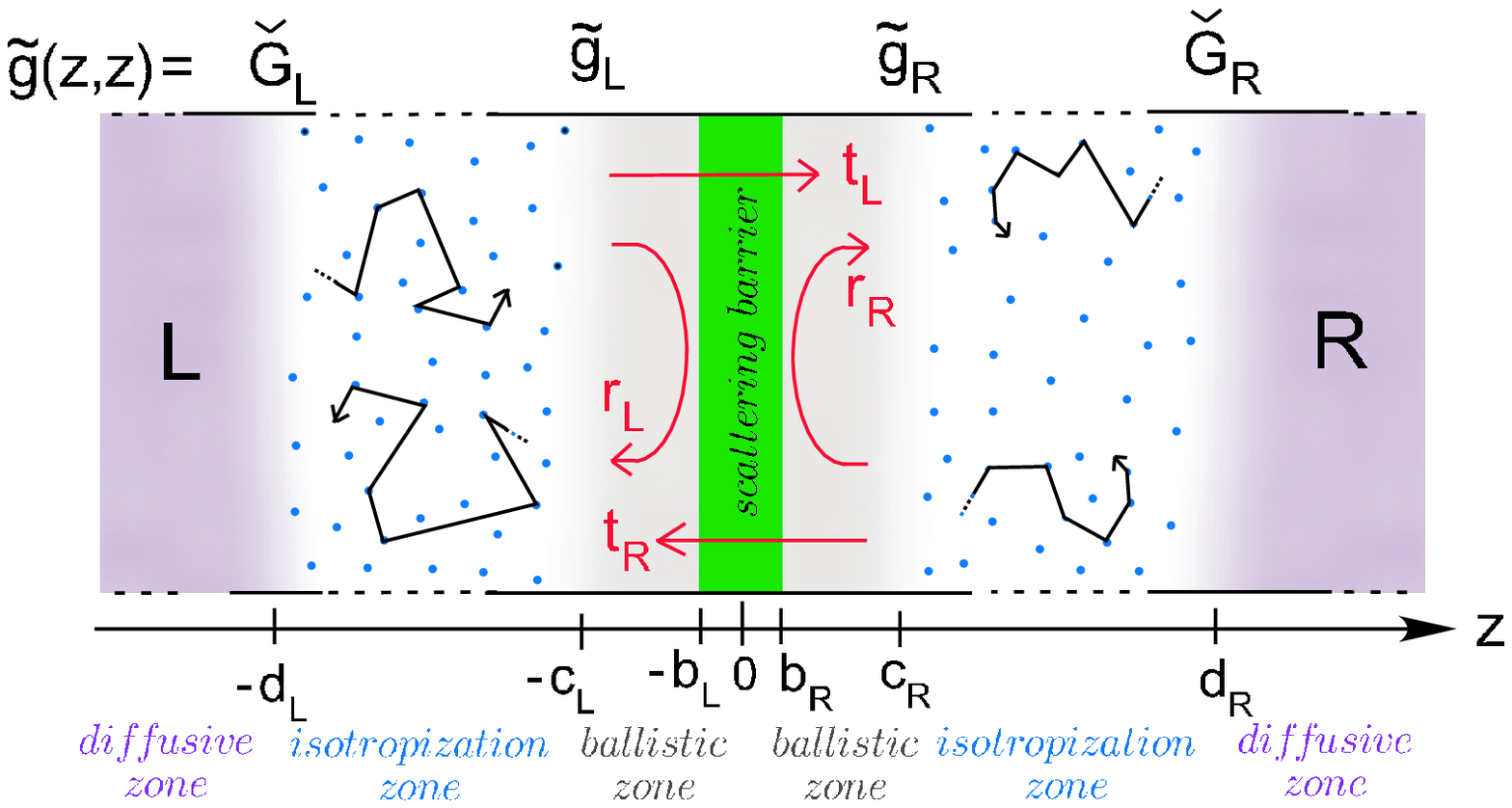}\label{Dessin}\end{figure}

From section \ref{general} to \ref{WeakFerro}, we consider a planar metallic
contact between two diffusive conductors noted $L$ (left conductor) and $R$
(right conductor) (see Fig.~\ref{Dessin}). The conductor $L$[$R$] can exhibit
spin and/or superconducting correlations, due to its superconducting order
parameter $\Delta$ or exchange field $E_{ex}$, or due to the proximity effect
with other conductors. For
the primary description of electronic correlations in $L$ and $R$, one can use
a general Green's function $\mathbb{G}$ defined in the Keldysh$\otimes
$Nambu$\otimes$Spin$\otimes$Coordinate space. In the stationary case,
$\mathbb{G}$ can be defined as
\begin{equation}
\mathbb{G}(\vec{r},\vec{r}^{\,\prime},\varepsilon)=\int\frac{dt}{\hbar}\text{
}\mathbb{G}(\vec{r},\vec{r}^{\,\prime},t-t^{\prime})\exp\{i\varepsilon
\frac{t-t^{\prime}}{\hbar}\}\label{defG1}%
\end{equation}
with\cite{def}
\begin{equation}
\mathbb{G}(\vec{r},\vec{r}^{\,\prime},t-t^{\prime})=\left[
\begin{array}
[c]{cc}%
\mathbb{G}^{r}(\vec{r},\vec{r}^{\,\prime},t-t^{\prime}) & \mathbb{G}^{K}%
(\vec{r},\vec{r}^{\,\prime},t-t^{\prime})\\
0 & \mathbb{G}^{a}(\vec{r},\vec{r}^{\,\prime},t-t^{\prime})
\end{array}
\right]  \label{defG2}%
\end{equation}%
\begin{equation}
\mathbb{G}^{r(a)}(\vec{r},\vec{r}^{\,\prime},t-t^{\prime})=\mp i\theta
(\pm(t-t^{\prime}))\check{\tau}_{3}\left\langle \left\{  \mathbf{\Psi}%
(t,\vec{r}),\mathbf{\Psi}^{\dagger}(t^{\prime},\vec{r}^{\,\prime})\right\}
\right\rangle \label{defG3}%
\end{equation}
and
\begin{equation}
\mathbb{G}^{K}(\vec{r},\vec{r}^{\,\prime},t-t^{\prime})=-i\check{\tau}%
_{3}\left\langle \left[  \mathbf{\Psi}(t,\vec{r}),\mathbf{\Psi}^{\dagger
}(t^{\prime},\vec{r}^{\,\prime})\right]  \right\rangle \label{DefG5}%
\end{equation}
Here, $\left[  ..,..\right]  $ and $\left\{  ..,..\right\}  $ denote
commutators and anticommutators respectively, $\vec{r}$, $\vec{r}^{\,\prime}$
space coordinates, $t$, $t^{\prime}$ time coordinates, and $\varepsilon$ the
energy\textbf{.} We use a spinor representation of the fermion operators,
i.e.
\begin{equation}
\mathbf{\Psi}^{\dagger}(t,\vec{r})=(\Psi_{\uparrow}^{\dagger}(t,\vec{r}%
),-\Psi_{\downarrow}^{\dagger}(t,\vec{r}),\Psi_{\uparrow}(t,\vec{r}%
),\Psi_{\downarrow}(t,\vec{r}))\label{phi}%
\end{equation}
in the Nambu$\otimes$Spin space. We denote by $\check{\tau}_{3}$ the third
Nambu Pauli matrix, i.e. $\check{\tau}_{3}=\mathrm{diag}(1,1,-1,-1)$ in the
Nambu$\otimes$Spin space. For later use, we also define the third spin Pauli
matrix i.e. $\check{\sigma}_{Z}=\mathrm{diag}(1,-1,1,-1)$. With the above
conventions, the Green's function $\mathbb{G}$ follows the Gorkov equations:
\begin{equation}
(\varepsilon\check{\tau}_{3}-H(\vec{r})+i\check{\Delta}(z)-\check{\Sigma
}_{imp}(z))\mathbb{G}(\vec{r},\vec{r}^{\,\prime},\varepsilon)=\delta(\vec
{r},\vec{r}^{\,\prime})\label{gorkov}%
\end{equation}
and%
\begin{equation}
\mathbb{G}(\vec{r},\vec{r}^{\,\prime},\varepsilon)(\varepsilon\check{\tau}%
_{3}-H(\vec{r}^{\,\prime})+i\check{\Delta}(z^{\prime})-\check{\Sigma}%
_{imp}(z^{\prime}))=\delta(\vec{r},\vec{r}^{\,\prime})\label{gorkov2}%
\end{equation}
Here, $\check{\Delta}$ corresponds to the gap matrix associated to a BCS
superconductor (see definition in Appendix \ref{UsadelAppendix}). The
Hamiltonian $H(\vec{r})$ can be decomposed as
\begin{equation}
H(\vec{r})=H_{l}(z)+H_{t}(\vec{\rho})+\bar{V}_{b}(z,\vec{\rho})\label{HH}%
\end{equation}
with $z$ and $\vec{\rho}$ the longitudinal and transverse components of
$\vec{r}$. The part $H_{l}(z)=-(\hbar^{2}/2m)\partial^{2}/\partial_{z}%
^{2}-E_{ex}(z)\check{\sigma}_{Z}-E_{F}(z)$ includes a ferromagnetic exchange
field $E_{ex}(z)$ in the direction $\vec{Z}$, and the Fermi energy
$E_{F}(z\lessgtr0)=E_{F,L(R)}$, whereas the part $H_{t}(\vec{\rho}%
)=-(\hbar^{2}/2m)\partial^{2}/\partial_{\vec{\rho}}^{2}+V_{c}(\vec{\rho})$
includes a lateral confinement potential $V_{c}(\vec{\rho})$. The potential
barrier $\bar{V}_{b}(z,\vec{\rho})$ describes a possibly spin-dependent and
non-specular interface. It is finite in the area $z\in\lbrack-b_{L},b_{R}]$
only. In the Born approximation, the impurity self-energy at side
$Q\in\{L,R\}$ of the interface can be expressed as $\check{\Sigma}%
_{imp}(z,\varepsilon)=-i\hbar\check{G}(z,\varepsilon)/2\tau_{Q}$. Here, the
impurity elastic scattering time $\tau_{Q}$ in material $Q$ can be considered
as spin-independent due to $E_{ex}\ll E_{F}$. The Green's function $\check
{G}(z,\varepsilon)$, which has already been mentioned in section
\ref{introduction}, corresponds to the quasiclassical and isotropic average of
$\mathbb{G}$ inside conductor $L(R)$. It can be calculated
as\cite{RefWolfgang1}
\begin{equation}
\check{G}(z,\varepsilon)=i\mathbb{G}(\vec{r}=\vec{R},\vec{r}^{\,\prime}%
=\vec{R},\varepsilon)/\pi\nu_{0}\label{qci}%
\end{equation}
with $z$ the longitudinal component of $\vec{R}$ and $\nu_{0}$ the density of
states per spin direction and unit volume for free electrons. Note that we
consider geometries where $\check{G}$, $\check{\Sigma}_{imp}$ and
$\check{\Delta}$ are independent of $\vec{\rho}$.

In this article, we consider the diffusive (i.e. quasiclassical and isotropic)
limit, i.e.
\begin{equation}
E_{ex},|\Delta|,\varepsilon,k_{B}T\ll\hbar/\tau_{Q}\ll E_{F} \label{QC}%
\end{equation}
where $T$ is the temperature and $k_{B}$ the Boltzmann constant. In this
regime, the spatial evolution of $\check{G}(z,\varepsilon)$ inside $L$ and $R$
is described by the Usadel equations which follow from Eqs.~(\ref{gorkov}) and
(\ref{gorkov2}) [see Appendix \ref{UsadelAppendix}]. The characteristic
distances occurring in the Usadel equations are $\xi_{F}=(\hbar D_{F}%
/E_{ex})^{1/2}$, $\xi_{N}=(\hbar D_{N}/2k_{B}T)^{1/2}$ and $\xi_{S}=(\hbar
D_{S}/2\Delta)^{1/2}$ for a ferromagnet $F$, a normal metal $N$ and a
superconductor $S$, respectively, with $D_{Q}=\upsilon_{F,Q}^{2}\tau_{Q}/3$
and $\upsilon_{F,Q}$ the diffusion constant and Fermi velocities of material
$Q$. According to Eq. (\ref{QC}), the scale $\xi_{Q}$ is much larger than the
elastic mean free path $\ell_{\text{e}}^{Q}=\upsilon_{F,Q}\tau_{Q}$.\textbf{
}Importantly, the Usadel equations alone are not sufficient to describe the
behavior of diffusive hybrid circuits. One also needs to relate the values of
$\check{G}$ at both sides of an $L/R$ interface with BCIGF, which we derive in
the next sections.

For the sake of concreteness, we give typical order of magnitudes for the
different lenghtscales involved in the problem. These lenghtscales strongly
depend on the detailed composition and structure of the materials and
interfaces considered, so that the applicability of the quasiclassical
isotropic description has to be checked in each case. The value of $b_{L(R)}$
can strongly vary from a few atomic layers to a few nanometers if the two
materials constituting the interface interdiffuse\cite{Robinson}. The mean
free path, which strongly depends on the impurity concentration, can be of the
order of a few nanometers\cite{Ryazanov}. The superconducting lenghtscale
$\xi_{S}$ is usually of the order of $10$~nm for Niobium\cite{Kim,Moraru}. The
Cooper pair penetration length can reach $\xi_{F}\sim10$~nm for a diluted
magnetic allow like CuNi\cite{Ryazanov}, or $\xi_{N}\sim1000$~nm for a normal
metal like Cu at $T=20$~mK\cite{gueron}.

It is worth to note, at this point, that the derivation presented below is not
restricted to stationary problems on superconducting heterostructures.
Actually most of the derivations made below do not rely on the specific
Keldysh structure introduced in Eqs.~(\ref{defG1})-(\ref{DefG5}) and our
results can be directly used to describe full counting statistics in the
extended Keldysh technique \cite{wn01} or multiple Andreev reflections
\cite{CuevasBelzig03}. In fact, boundary conditions for arbitrary
time-dependent scattering problems have been recently formulated in a similar
spirit \cite{SnymanNazarov08}. However, having in mind the many concrete
applications of the boundary conditions in superconducting heterostructures
and keeping the notation as simple as possible, we derive the BCIGF below in
the framework of the stationary Keldysh-Nambu Greens functions.

\section{Ballistic Green's function \label{ballistic2}}

Considering the structure of Eqs.~(\ref{gorkov})-(\ref{HH}), for $z,z^{\prime
}<-b_{L}$ or $z,z^{\prime}>b_{R}$, one can expand $\mathbb{G}$ in transverse
modes as\cite{RefZaitsev}
\begin{align}
&  \mathbb{G}^{\nu\sigma,\nu^{\prime}\sigma^{\prime}}(\vec{r},\vec
{r}^{\,\prime},\varepsilon)\nonumber\\
&  =\sum_{ns,n^{\prime}s^{\prime}}\left(  \widetilde{\mathcal{G}%
}_{ns,n^{\prime}s^{\prime}}^{\nu\sigma,\nu^{\prime}\sigma^{\prime}%
}(z,z^{\prime},\varepsilon)\frac{\chi_{n}(\vec{\rho})\chi_{m}^{\ast}(\vec
{\rho}^{\prime})}{2\pi\hbar\sqrt{\upsilon_{n}(z,\varepsilon)\upsilon
_{m}(z^{\prime},\varepsilon)}}\right. \nonumber\\
&  \qquad\qquad\left.  \times\exp\left[  isk_{n}(z)z-is^{\prime}%
k_{m}(z^{\prime})z^{\prime}\right]  \right)  \label{Zaitsev}%
\end{align}
In this section, we use spin indices $\sigma,\sigma^{\prime}\in\{\uparrow
,\downarrow\}$ which correspond to spin directions parallel or antiparallel to
the direction $\vec{Z}$, and Nambu indices $\nu,\nu^{\prime}\in\{e,h\}$ for
electron and hole states. The indices $s,s^{\prime}\in\{+,-\}$ account for the
longitudinal direction of propagation (we use $\sigma,\sigma^{\prime}=\pm1$,
$\nu,\nu^{\prime}=\pm1$, and $s,s^{\prime}=\pm1$ in mathematical expressions).
We introduce the wavefunction $\chi_{n}(\vec{\rho})$ for the transverse
channel $n$, i.e. $H_{t}(\vec{\rho})\chi_{n}(\vec{\rho})=E_{n}\chi_{n}%
(\vec{\rho})$, and the corresponding longitudinal momentum and velocity, i.e.
$k_{n}(z)=(2m(E_{F}(z)-E_{n}))^{1/2}/\hbar$ and $\upsilon_{n}(z)=\hbar
k_{n}(z)/m$. Importantly, we have disregarded the dependences of $k_{n}$ and
$\upsilon_{n}$ on $E_{ex}$ and $\varepsilon$ due to Eq.~\ref{QC}. The
$\widetilde{~}$ decoration denotes that the Green's function $\widetilde
{\mathcal{G}}$ can have a general structure in the Keldysh$\otimes
$Nambu$\otimes$Spin$\otimes$Channel$\otimes$Direction space, noted
$\mathcal{E}$ in the following. In contrast, $\check{}$ denotes the fact that
$\check{G}_{L(R)}$ has no structure in the Channel$\otimes$Direction
sub-space, noted $\underline{\mathcal{E}}$ in the following (see the summary
of notations in Table \ref{decoration}). \begin{table}[tb]
\quad\quad Structure allowed in the subspaces of:\newline%
\begin{tabular}
[c]{|c|c|c|c|c|c|}\hline
\quad\quad\quad & channels $n$ & direction $s$ & spins $\sigma$ & Nambu $\nu$
& Keldysh\\\hline
$\widetilde{A}$ & yes & yes & yes & yes & yes\\\hline
$\check{A}$ & no & no & yes & yes & yes\\\hline
$\breve{A}$ & yes & yes & yes & diagonal & no\\\hline
$\bar{A}$ & yes & yes & yes & no & no\\\hline
$\hat{A}$ & yes & yes & no & no & no\\\hline
$\overline{\overline{A}}$ & yes & no & no & no & no\\\hline
\end{tabular}
\caption{Meaning of the various decorations used in this article for operators
defined in the $\mathcal{E}$ space. Each decoration corresponds to a
particular structure in the transverse channels (index $n$), propagation
direction (index $s$), spin (index $\sigma$), Nambu (index $\nu$) and Keldysh
subspaces.}%
\label{decoration}%
\end{table}Due to Eqs.~(\ref{gorkov}) and (\ref{gorkov2}), $\widetilde
{\mathcal{G}}(z,z^{\prime},\varepsilon)$ is not continuous at $z=z^{\prime}%
$\cite{RefZaitsev,RefYuli}. One can use\cite{Shelankov,DefSign}
\begin{equation}
\widetilde{\mathcal{G}}(z,z^{\prime},\varepsilon)=-i\pi\left(  \tilde
{g}(z,z^{\prime},\varepsilon)+\hat{\Sigma}_{3}\text{\textrm{sign}}%
(z-z^{\prime})\right)  \label{match}%
\end{equation}
with $\hat{\Sigma}_{3}$ the third Pauli matrix in the direction of propagation
space, i.e. $(\hat{\Sigma}_{3})_{ns,ms^{\prime}}^{\sigma,\sigma^{\prime}%
}=s\delta_{ss^{\prime}}\delta_{nm}\delta_{\sigma\sigma^{\prime}}\delta_{\nu
\nu^{\prime}}\mathbf{1}_{K}$. Equation~(\ref{match}) involves a
\emph{ballistic Green's function} $\tilde{g}(z,z^{\prime},\varepsilon
)\in\mathcal{E}$ which is continuous at $z=z^{\prime}$. We will see below that
this quantity plays a major role in the derivation of the BCIGF.

For later use, we now derive the equations of evolution followed by
$\widetilde{\mathcal{G}}$. Inserting Eq.~(\ref{Zaitsev}) into
Eqs.~(\ref{gorkov}-\ref{gorkov2}), one can check that, for $z\neq z^{\prime}$
and $z,z^{\prime}<-b_{L}$ ($Q=L$) or $z,z^{\prime}>b_{R}$ ($Q=R$),
$\widetilde{\mathcal{G}}$ follows the equations
\begin{equation}
\left[  i\hbar\hat{\Sigma}_{3}\overline{\overline{\upsilon}}_{Q}\frac
{\partial}{\partial z}+i\check{\Delta}-\check{\Sigma}_{imp}(z)\right]
\otimes\widetilde{\mathcal{G}}(z,z^{\prime},\varepsilon)=0 \label{elen1}%
\end{equation}
and
\begin{equation}
\widetilde{\mathcal{G}}(z,z^{\prime},\varepsilon)\otimes\left[  -i\hbar
\hat{\Sigma}_{3}\overline{\overline{\upsilon}}_{Q}\frac{\partial}{\partial
z^{\prime}}+i\check{\Delta}-\check{\Sigma}_{imp}(z^{\prime})\right]  =0\,.
\label{elen2}%
\end{equation}
We have introduced above a velocity matrix $\overline{\overline{\upsilon}}%
_{Q}$ with a structure in the channels subspace only, i.e. $(\overline
{\overline{\upsilon}}_{L(R)})_{ns,n^{\prime}s^{\prime}}^{\nu\sigma,\text{ }%
\nu^{\prime}\sigma^{\prime}}=\upsilon_{n}(z\lessgtr0)\delta_{ss^{\prime}%
}\delta_{nn^{\prime}}\delta_{\sigma\sigma^{\prime}}\delta_{\nu\nu^{\prime}%
}\mathbf{1}_{K}$, with $\mathbf{1}_{K}$ the identity matrix in the Keldysh
space. We have furthermore assumed that the so-called envelope function
$\widetilde{\mathcal{G}}$ varies smoothly on the scale of the Fermi wave
length, in order to neglect terms proportional to $\partial^{2}/\partial
z^{2}$ and $\partial^{2}/\partial z^{\prime2}$ in Eqs. (\ref{elen1}%
-\ref{elen2}) \cite{RefZaitsev}.

\section{Scattering description of a metallic contact \label{ballistic}}

We now define, at both sides of the barrier $\bar{V}_{b}$, two ballistic zones
(with no impurity scattering) located at $z\in\lbrack-c_{L},-b_{L}]$ and
$z\in\lbrack c_{R},b_{R}]$, with $c_{L(R)}-b_{L(R)}\lesssim\ell_{\text{e}%
}^{L(R)}$ (grey areas in Fig.~\ref{Dessin}). In the region $z\in\left[
-c_{L},c_{R}\right]  $, we can disregard the superconducting gap matrix
$\check{\Delta}$ since $c_{L}+c_{R}\ll\xi_{S}$. Therefore, the electron and
hole dynamics can be described with the Schr\"{o}dinger equation
\begin{equation}
\left[  \varepsilon\check{\tau}_{3}-H(\vec{r})\right]  \phi(\vec
{r},\varepsilon)=0 \label{l1}%
\end{equation}
or, equivalently,%
\begin{equation}
\phi^{\dagger}(\vec{r},\varepsilon)\left[  \varepsilon\check{\tau}_{3}%
-H(\vec{r})\right]  =0 \label{l2}%
\end{equation}
whose solution has the form\cite{BlanterButtiker}%
\begin{equation}
\phi^{\nu,\sigma}(\vec{r},\varepsilon)=\sum\limits_{n,s}\psi_{n,s}^{\nu
,\sigma}(z,\varepsilon)\frac{\chi_{n}(\vec{\rho})}{\sqrt{2\pi\hbar\upsilon
_{n}(z)}}e^{is\nu k_{n}(z)z} \label{phiphi}%
\end{equation}
in the ballistic zones. Here, $\phi(\vec{r},\varepsilon)$ is a vector in the
Spin$\otimes$Nambu$\otimes$Keldysh space, and $\psi(z,\varepsilon)$ is a
vector in the $\mathcal{E}$ space. The index $s$ corresponds again to the
longitudinal direction of propagation. We have introduced $\nu$ indices in the
exponential factors of Eq.~(\ref{phiphi}) because, for the same sign of
wavevector, electrons and holes go in opposite directions. Therefore, in
Eq.~(\ref{phiphi}), $s=+/-1$ systematically denotes the right/left going
states. One can introduce a transfer matrix $\mathcal{\breve{M}}%
(\varepsilon)\in\mathcal{E}$ such that $\psi(c_{R},\varepsilon
)=\mathcal{\breve{M}}(\varepsilon)\psi(-c_{L},\varepsilon)$. The matrix
$\mathcal{\breve{M}}$ and the Landauer-B\"{u}ttiker scattering matrix can be
considered as equivalent descriptions of a contact, provided one introduces
small but finite transmission coefficients to regularize $\mathcal{\breve{M}}$
in case of perfectly reflecting channels. This regularization procedure does
not affect practical calculations as illustrated in Section VII.D. Since
$H(\vec{r})$ does not couple electron and holes, $\mathcal{\breve{M}}$ has the
structure
\begin{equation}
\mathcal{\breve{M}}=\left[
\begin{array}
[c]{cc}%
\mathcal{M}^{e}(\varepsilon) & 0\\
0 & \mathcal{M}^{e}(-\varepsilon)^{\ast}%
\end{array}
\right]
\end{equation}
in the Nambu subspace. Moreover, $\mathcal{\breve{M}}$ is proportional to the
identity in the Keldysh space, like $H(\vec{r})$. For later use, we point out
that flux conservation leads to\cite{RefStone}
\begin{equation}
\mathcal{\breve{M}}^{\dagger}\text{ }\hat{\Sigma}_{3}\text{ }\mathcal{\breve
{M}}=\mathcal{\breve{M}}\text{ }\hat{\Sigma}_{3}\text{ }\mathcal{\breve{M}%
}^{\dagger}=\hat{\Sigma}_{3} \label{fluxConservation}%
\end{equation}

We now connect the above scattering approach with the Green's function
description\cite{Millis}. With the assumptions done in this section,
Eqs.~(\ref{gorkov}-\ref{gorkov2}) give, for $z\in\left[  -c_{L},c_{R}\right]
$ and $z^{\prime}\neq z$%

\begin{equation}
(\varepsilon\check{\tau}_{3}-H(\vec{r}))\mathbb{G}(\vec{r},\vec{r}^{\,\prime
},\varepsilon)=0 \label{l3}%
\end{equation}
and%
\begin{equation}
\mathbb{G}(\vec{r},\vec{r}^{\,\prime},\varepsilon)(\varepsilon\check{\tau}%
_{3}-H(\vec{r}^{\,\prime}))=0 \label{l4}%
\end{equation}
We recall that in the ballistic zones, $\mathbb{G}$ takes the form
(\ref{Zaitsev}). In the domain $z^{\prime}>z$, a comparison between
Eqs.~(\ref{l3}-\ref{l4}) and (\ref{l1}-\ref{l2}) gives, in terms of the
decompositions (\ref{Zaitsev}) and (\ref{phiphi})
\begin{align*}
&  \widetilde{U}\widetilde{\mathcal{G}}(c_{R},c_{R}+0^{-},\varepsilon
)\widetilde{U}\\
&  =\mathcal{\breve{M}}(\varepsilon)\widetilde{U}\widetilde{\mathcal{G}%
}(-c_{L}+0^{+},-c_{L},\varepsilon)\widetilde{U}\mathcal{\breve{M}}^{\dagger
}(\varepsilon)
\end{align*}
We have introduced above the transformation $\widetilde{U}=(\mathbf{1}%
+\check{\tau}_{3}+[\mathbf{1}-\check{\tau}_{3}]\hat{\Sigma}_{1})/2$ to
compensate the fact that the $\nu$ indices do not occur in the exponential
terms of Eq.~(\ref{Zaitsev}). Using Eq.~(\ref{match}), we obtain
\begin{equation}
\tilde{g}_{R}=\bar{M}\tilde{g}_{L}\bar{M}^{\dag} \label{rel}%
\end{equation}
with $\tilde{g}_{L}=\tilde{g}(-c_{L},-c_{L},\varepsilon)$, $\tilde{g}%
_{R}=\tilde{g}(c_{R},c_{R},\varepsilon)$, and%
\begin{equation}
\bar{M}=\left[
\begin{array}
[c]{cc}%
\mathcal{M}^{e}(\varepsilon) & 0\\
0 & \hat{\Sigma}_{1}\mathcal{M}^{e}(-\varepsilon)^{\ast}\hat{\Sigma}_{1}%
\end{array}
\right]  \label{Mbar}%
\end{equation}
in the Nambu subspace. Note that due to Eq.~(\ref{fluxConservation}), one has%
\begin{equation}
\bar{M}^{\dag}\text{ }\hat{\Sigma}_{3}\text{ }\bar{M}=\bar{M}\text{ }%
\hat{\Sigma}_{3}\text{ }\bar{M}^{\dag}=\hat{\Sigma}_{3}\,.
\label{FluxConservation}%
\end{equation}

We now discuss how spin-dependences arise in our problem. Due to the
hypotheses required to reach the diffusive limit [see Eq.~(\ref{QC})], we have
neglected the dependence of $k_{n}$ and $\upsilon_{n}$ on the exchange field
$E_{ex}$ and the energy $\varepsilon$. Accordingly, we have to disregard the
dependence of $\bar{M}$ on $E_{ex}$ and $\varepsilon$. This does not forbid
that $\bar{M}$ depends on spin. Indeed, in the general case, when an interface
involves a $F$ material which is ferromagnetic in the bulk, the transfer
matrix $\bar{M}$ can depend on spin for two reasons: first, the wavectors of
the electrons scattered by the barrier can depend on spin due to $E_{ex}$, and
second, the interface barrier potential $\bar{V}_{b}$ can itself be
spin-dependent. Importantly, one can check that $E_{ex}$ and $\bar{V}_{b}$
occur independently in Eqs. (\ref{gorkov}-\ref{gorkov2}). The value of
$E_{ex}$ and the spin-dependence of $\bar{V}_{b}$ are not directly related,
because the second depend on properties like interfacial disorder or
discontinuities in the electronic band structure, which do not influence
$E_{ex}$ far from the interface. Therefore, nothing forbids to have
simultaneously $E_{ex}\ll E_{F}$ (this can occur e.g. in a diluted
ferromagnetic alloy like PdNi) and a spin-dependent $\bar{M}$, due to a
spin-dependent interface potential $\bar{V}_{b}$. It is even possible to
obtain this situation artificially, by fabricating e.g. a contact with a very
thin $FI$ barrier separating two normal metals or superconductors. Note that
in spite of $E_{ex}\ll E_{F}$, the exchange field $E_{ex}$ can play a major
role in diffusive hybrid circuits by modifying drastically the spatial
evolution of the isotropic Green's function $\check{G}_{F}(z,\varepsilon)$
inside a ferromagnetic metal $F$ on the scale $\xi_{F}$ [see
Appendix~\ref{UsadelAppendix}].

\section{Isotropization scheme\label{isotropization}}

In this section, we show that the Green's function $\tilde{g}(z,z^{\prime
}=z,\varepsilon)$ becomes isotropic in momentum space (i.e. proportional to
the identity in the \underline{$\mathcal{E}$} subspace) due to impurity
scattering, when moving further away from the contact. One can consider that
this process occurs in "isotropization zones" with a size $d_{L(R)}$ of the
order of a few $\ell_{\text{e}}^{L(R)}$ for side $L(R)$ of the
contact\cite{scale} (dotted areas in Fig.~\ref{Dessin}). Beyond the
isotropization zones, quasiparticles reach diffusive zones (purple areas in
Fig.~\ref{Dessin}) characterized by isotropic Green's functions $\check
{G}(z,\varepsilon)$ with no structure in the \underline{$\mathcal{E}$}
subspace. We show below that $\tilde{g}(z,z^{\prime}=z,\varepsilon)$ tends to
$\check{G}(z=\mp d_{L(R)},\varepsilon)$ at the external borders $z=\mp
d_{L(R)}$ of the isotropization zones. Note that the results presented in this
section do not depend on the details of the isotropization mechanism.

We study the spatial evolution of $\widetilde{\mathcal{G}}$ in the
isotropization zones located at $z\in\lbrack-d_{L},-c_{L}]$ and $z\in\lbrack
c_{R},d_{R}]$, using Eqs.~(\ref{elen1}) and (\ref{elen2}). The superconducting
gap matrix $\check{\Delta}$ can be neglected from these Eqs. due to
$d_{L(R)}\ll\xi_{S}$. We thus obtain, for the isotropization zone of side $Q$
and $z\neq z^{\prime}$
\begin{equation}
\left(  \hat{\Sigma}_{3}\overline{\overline{\upsilon}}_{Q}\frac{\partial
}{\partial{z}}+\frac{\check{G}(z,\varepsilon)}{2\tau_{Q}}\right)
\otimes\widetilde{\mathcal{G}}(z,z^{\prime},\varepsilon)=0 \label{elen3}%
\end{equation}
and%
\begin{equation}
\widetilde{\mathcal{G}}(z,z^{\prime},\varepsilon)\otimes\left(  -\hat{\Sigma
}_{3}\overline{\overline{\upsilon}}_{Q}\frac{\partial}{\partial{z^{\prime}}%
}+\frac{\check{G}(z^{\prime},\varepsilon)}{2\tau_{Q}}\right)  =0 \label{elen4}%
\end{equation}
Due to $\xi_{Q}\gg\ell_{\text{e}}^{Q}$, one can disregard the space-dependence
of $\check{G}(z,\varepsilon)$ in the above equations. We will thus replace
$\check{G}(z,\varepsilon)$ by its value $\check{G}_{Q}$ at the beginning of
the diffusive zone $Q$, i.e. $\check{G}_{L(R)}=\check{G}(z=\mp d_{L(R)}%
,\varepsilon)$. For later use, we recall that $\check{G}_{L}$ and $\check
{G}_{R}$ fulfill the normalization condition
\begin{equation}
\check{G}_{L}^{2}=\check{G}_{R}^{2}=\mathbf{1} \label{normalized}%
\end{equation}
with $\mathbf{1}$ the identity in the $\mathcal{E}$ space. In the
isotropization zone of side $Q$, Eqs.~(\ref{match}) and (\ref{elen3}%
-\ref{normalized}) give
\begin{align}
\widetilde{\mathcal{G}}(z,z^{\prime},\varepsilon)  &  =-i\pi\tilde{P}%
_{Q}[\lambda_{Q}(z)]\label{Gspace}\\
&  \times\left[  \tilde{g}_{Q}+\text{\textrm{sign}}(z-z^{\prime})\hat{\Sigma
}_{3}\right]  \tilde{P}_{Q}[-\lambda_{Q}(z^{\prime})]\nonumber
\end{align}
with $\lambda_{L(R)}(z)=z\pm c_{L(R)}$ and
\begin{equation}
\widetilde{P}_{Q}[z]=\mathrm{ch}\left[  z/2\overline{\overline{\upsilon}}%
_{Q}\tau_{Q}\right]  -\hat{\Sigma}_{3}\check{G}_{Q}\mathrm{sh}\left[
z/2\overline{\overline{\upsilon}}_{Q}\tau_{Q}\right]  \label{Gspace2}%
\end{equation}
for $Q\in\{L,R\}$. Note that the choice of the coordinate $d_{L(R)}$ in Fig.1
is somewhat arbitrary, i.e. defined only up to an uncertainty of the order of
$\ell_{\text{e}}^{L(R)}$, because there is a smooth transition between the
isotropization and diffusive zones of the contact. As a result, $\widetilde
{\mathcal{G}}$ must tend continuously to its limit value $\widetilde
{\mathcal{G}}_{diff}$ in the diffusive zones. The function $\widetilde
{\mathcal{G}}_{diff}(z,z^{\prime},\varepsilon)$ must vanish for $\left\vert
z-z^{\prime}\right\vert \gg\ell_{\text{e}}^{Q}$ (see e.g. Ref.
\onlinecite{Abrikosov}). This imposes to cancel the "exponentially divergent"
terms in Eq. (\ref{Gspace}) , which requires\cite{RefYuli}
\begin{equation}
\left(  \hat{\Sigma}_{3}+\check{G}_{L}\right)  \left(  \tilde{g}_{L}%
-\hat{\Sigma}_{3}\right)  =0 \label{Iso1}%
\end{equation}%
\begin{equation}
\left(  \tilde{g}_{L}+\hat{\Sigma}_{3}\right)  \left(  \hat{\Sigma}_{3}%
-\check{G}_{L}\right)  =0\,, \label{Iso2}%
\end{equation}%
\begin{equation}
\left(  \hat{\Sigma}_{3}-\check{G}_{R}\right)  \left(  \tilde{g}_{R}%
+\hat{\Sigma}_{3}\right)  =0\,, \label{Iso3}%
\end{equation}%
\begin{equation}
\left(  \tilde{g}_{R}-\hat{\Sigma}_{3}\right)  \left(  \hat{\Sigma}_{3}%
+\check{G}_{R}\right)  =0\,. \label{Iso4}%
\end{equation}
For $z\rightarrow\mp d_{L(R)}$ we obtain from Eqs.~(\ref{Gspace})-(\ref{Iso4})
that $\widetilde{\mathcal{G}}$ finally approaches
\begin{align}
\widetilde{\mathcal{G}}_{diff}(z,z^{\prime},\varepsilon)  &  =-i\pi\exp
(-\frac{\left\vert z-z^{\prime}\right\vert }{2\overline{\overline{\upsilon}%
}_{L(R)}\tau_{L(R)}})\nonumber\\
&  \times\left(  \check{G}_{L(R)}+\text{\textrm{sign}}(z-z^{\prime}%
)\hat{\Sigma}_{3}\right)  \,, \label{limit}%
\end{align}
so that $\tilde{g}_{L(R)}(z,z^{\prime}=z,\varepsilon)$ tends to $\check
{G}_{L(R)}$. As required, the expression (\ref{limit}) of $\widetilde
{\mathcal{G}}_{diff}$ does not depend on the exact choice of the coordinate
$d_{L(R)}$ and vanishes for $\left\vert z-z^{\prime}\right\vert \gg
\ell_{\text{e}}^{Q}$. Equations~(\ref{Gspace}-\ref{Iso4}) indicate that the
decay length for the isotropization of $\tilde{g}_{L(R)}(z,z^{\prime
}=z,\varepsilon)$ is $\max_{n}[(2m[E_{F,L(R)}-E_{n}])^{1/2}\tau_{L(R)}%
]=\ell_{\text{e}}^{L(R)}$, as anticipated above. Moreover, inserting
Eq.~(\ref{limit}) into Eq.~(\ref{Zaitsev}) leads to an expression of
$\mathbb{G}$ whose semiclassical and isotropic average corresponds to
$\check{G}_{L(R)}$, as expected\cite{QCaverage}. Importantly, from Eqs.
(\ref{Gspace}-\ref{Iso4}), one sees explicitly that $\widetilde{\mathcal{G}}$
is smooth on a scale of the Fermi wave length, which justifies a posteriori
the use of the approximated Eqs. (\ref{elen1}) and (\ref{elen2}) in this
section\textbf{.}

\section{Matrix current and general boundary conditions\label{generalBC}}

Our purpose is to establish a relation between $\check{G}_{L}$ and $\check
{G}_{R}$. To complete this task, it is convenient to introduce the matrix
current\cite{RefYuli}
\begin{equation}
\check{I}(z,\varepsilon)=\frac{\mathrm{e}^{2}\hbar}{\pi m}\int d\rho\left.
\left(  \frac{\partial}{\partial z}-\frac{\partial}{\partial z^{\prime}%
}\right)  \mathbb{G}(\vec{r},\vec{r}^{\text{ }\prime},\varepsilon)\right\vert
_{\vec{r}\text{ }=\text{ }\vec{r}^{\text{ }\prime}}\,. \label{current1}%
\end{equation}
This quantity characterizes the transport properties of the circuit for
coordinate $z$ and energy $\varepsilon$. It contains information on the charge
current (see section \ref{chargeCurrent}) but also on the flows of spins and
electron-hole coherence. Note that in this article, $\mathrm{e}$ denotes the
absolute value of the electron charge. Using Eq.~(\ref{Zaitsev}) and the
orthonormalization of the transverse wave functions $\chi_{n}^{\sigma}$, the
matrix current is written as
\begin{equation}
\check{I}(z,\varepsilon)=2iG_{q}\mathrm{Tr}_{n,s}\left[  \hat{\Sigma}%
_{3}\widetilde{\mathcal{G}}(z,z,\varepsilon)\right]  /\pi\,. \label{current2N}%
\end{equation}
for $z<-b_{L}$ or $z>b_{R}$. Here $\mathrm{Tr}_{n,s}$ denotes the trace in the
\underline{$\mathcal{E}$} sub-space and $G_{q}\equiv\mathrm{e}^{2}/2\pi\hbar$
is the conductance quantum. Inside the isotropization zones, using
Eq.~(\ref{Gspace}), one obtains\cite{DefSign}
\begin{equation}
\check{I}(z,\varepsilon)=2G_{q}\mathrm{Tr}_{n,s}\left[  \hat{\Sigma}%
_{3}\widetilde{P}_{Q}[\lambda_{Q}(z)]\tilde{g}_{Q}\widetilde{P}_{Q}%
[-\lambda_{Q}(z)]\right]  \,.
\end{equation}
Considering that $\widetilde{P}_{Q}(z)$ has a structure in the \underline
{$\mathcal{E}$} sub-space only, and that $\widetilde{P}_{Q}[-\lambda
_{Q}(z)]\hat{\Sigma}_{3}\widetilde{P}_{Q}[\lambda_{Q}(z)]=\hat{\Sigma}_{3}$,
one finds
\begin{equation}
\check{I}(z,\varepsilon)=2G_{q}\mathrm{Tr}_{n,s}\left[  \hat{\Sigma}_{3}%
\tilde{g}_{L(R)}\right]  =\check{I}_{L(R)}(\varepsilon) \label{Iiso}%
\end{equation}
at any point in the left(right) isotropization zone. We conclude that, quite
generally, the matrix current is conserved inside each isotropization zone. We
will see in next paragraph that this property is crucial to derive the BCIGF.

In order to express $\tilde{g}_{L}$ in terms of $\check{G}_{L}$ and $\check
{G}_{R}$ and $\bar{M}$, we multiply Eq.~(\ref{Iso1}) by $\check{G}_{L}$ from
the left and Eq.~(\ref{Iso3}) by $\check{G}_{L}\bar{M}^{\dag}$ from the left
and by $(\bar{M}^{\dag})^{-1}$ from the right. Then, we add up the two
resulting equations after simplifications based on Eqs.~(\ref{rel}),
(\ref{FluxConservation}), and (\ref{normalized}). This leads to
\begin{equation}
\check{I}_{L}(\varepsilon)=2G_{q}\mathrm{Tr}_{n,s}\left[  2\widetilde{D}%
_{L}^{-1}\left(  \check{G}_{L}\hat{\Sigma}_{3}+\mathbf{1}\right)
-\mathbf{1}\right]  \label{GBC}%
\end{equation}
with $\widetilde{D}_{L}=\mathbf{1}+\check{G}_{L}\bar{M}^{\dag}\check{G}%
_{R}\bar{M}$. A similar calculation leads to%
\begin{equation}
\check{I}_{R}(\varepsilon)=2G_{q}\mathrm{Tr}_{n,s}\left[  2\widetilde{D}%
_{R}^{-1}\left(  \check{G}_{R}\hat{\Sigma}_{3}-\mathbf{1}\right)
+\mathbf{1}\right]  \label{GBC3}%
\end{equation}
with $\widetilde{D}_{R}=\mathbf{1}+\check{G}_{R}(\bar{M}^{\dag})^{-1}\check
{G}_{L}\bar{M}^{-1}$. Equations (\ref{GBC}) and (\ref{GBC3}) represent the
most general expression for $\check{I}_{L(R)}(\varepsilon)$ in terms of the
isotropic Green's functions $\check{G}_{L(R)}$ and the transfer matrix
$\bar{M}$. The conservation of the matrix current up to the beginning $z=\mp
d_{L(R)}$ of the diffusive zones allows to identify these expressions with
\begin{equation}
\check{I}_{L(R)}(\varepsilon)=-\frac{A}{\rho_{L[R]}}\left.  \check
{G}(z,\varepsilon)\frac{\partial\check{G}(z,\varepsilon)}{\partial
z}\right\vert _{z=\mp d_{L(R)}} \label{der}%
\end{equation}
Here, $\rho_{L(R)}$ denotes the resistivity of conductor $L(R)$ and $A$ the
junction area. Formally speaking, Eqs.~(\ref{GBC}), (\ref{GBC3}) and
(\ref{der}) complete our task of finding the general BCIGF for spin-dependent
and diffusive metallic interfaces. We recall that to derive these equations,
we have assumed a weak exchange field in ferromagnets ($E_{ex}\ll E_{F}$), as
required to reach the diffusive limit [see Eq. (\ref{QC})]. However, we have
made no restriction on the structure of the contact transfer matrix $\bar{M}$.
In particular, $\bar{M}$ can be arbitrarily spin-polarized, and it is not
necessarily spin-conserving or channel-conserving. However, at this stage, a
concrete calculation requires the knowledge of the full $\bar{M}$ (or
equivalently the full scattering matrix). Usually this information is not
available for realistic interfaces and one has to reduce Eqs.~(\ref{GBC}%
)-(\ref{GBC3}) to simple expressions, using some simplifying assumptions. For
a spin-independent tunnel interface, Eqs.~(\ref{GBC})-(\ref{GBC3}) can be
expressed in terms of the contact tunnel conductance $G_{T}$ only, which is a
formidable simplification\cite{Kuprianov}. Another possibility is to disregard
superconducting correlations. In this case, Eq.~(\ref{GBC}) and (\ref{GBC3})
lead to the normal-state BCIGF introduced in
Refs.~\onlinecite {RefCircuit,Braatas} (see appendix \ref{NormalState} for
details). The normal-state BCIGF involve the conductance $G_{T}$ but also a
coefficient $G_{MR}$ which accounts for the spin-dependence of the contact
scattering probabilities, and the transmission and reflection mixing
conductances $G_{mix}^{t}$ and $G_{mix}^{L(R),r}$ which account for
spin-torque effects and interfacial effective fields\cite{revuesFN}. We will
show below that for a circuit enclosing superconducting elements, the BCIGF
can also be simplified in various limits.

Note that since the transition between the ballistic, isotropization and
diffusive zones is smooth, the choice of the coordinates $d_{L(R)}$ and
$c_{L(R)}$ in Fig.1 is somewhat arbitrary, i.e. defined only up to an
uncertainty of the order of $\ell_{\text{e}}^{L(R)}$ or a fraction of
$\ell_{\text{e}}^{L(R)}$ respectively. However, one can check that this choice
does not affect the BCIGF. First, a change of $c_{L}$ and $c_{R}$ by
quantities $\delta c_{L}$ and $\delta c_{R}$ of the order of a fraction of
$\ell_{\text{e}}^{L(R)}$ requires to replace the matrix $\bar{M}$ appearing in
Eqs.~(\ref{GBC}-\ref{der}) by $\overline{\overline{A}}_{R}~\bar{M}%
~\overline{\overline{A}}_{L}$, where the matrices $\overline{\overline{A}}%
_{R}$ and $\overline{\overline{A}}_{L}$ have a non-trivial (i.e. diagonal)
structure in the \underline{$\mathcal{E}$} subspace only, with diagonal
elements $\overline{\overline{A}}_{L,n,s}=\exp[i~s~\delta c_{L}~k_{n}]$ and
$\overline{\overline{A}}_{R,n,s}=\exp[i~s~\delta c_{R}~k_{n}]$. Since
$\check{G}_{L(R)}$ commutes with $\overline{\overline{A}}_{R[L]}$, this leaves
the BCIGF unchanged. Second, due to Eqs. (\ref{Iso1}-\ref{Iso4}), the BCIGF do
not depend either on the exact values of $d_{L}$ and $d_{R}$.

\section{Case of a weakly spin-dependent $S/F$ contact\label{WeakFerro}}

\subsection{Perturbation scheme\label{Perturbation scheme}}

In the next sections, we assume that the transverse channel index $n$ and the
spin index $\sigma=\uparrow,\downarrow$ corresponding to spin components along
$\vec{Z}$ are conserved when electrons are scattered by the potential barrier
$\bar{V}_{b}$ between the two ballistic zones (we use for instance $\bar
{V}_{b}(z,\vec{\rho})=V_{0}(z)\check{\sigma}_{0}+V_{1}(z)\check{\sigma}_{Z}$).
In this case, one can describe the scattering properties of the barrier with
parameters $T_{n}$, $P_{n}$, $\varphi_{n}^{L(R)}$, and $d\varphi_{n}^{L(R)}$
defined from%

\begin{equation}
\left\vert t_{L(R),n\sigma}\right\vert ^{2}=T_{n}(1+\sigma P_{n}) \label{p1}%
\end{equation}
and%
\begin{equation}
\arg(r_{L(R),n\sigma})=\varphi_{n}^{L(R)}+\sigma(d\varphi_{n}^{L(R)}/2)
\label{p2}%
\end{equation}
with $t_{L(R),n\sigma}$ the transmission amplitude from side $L(R)$ to side
$R(L)$ of the barrier and $r_{L(R),n\sigma}$ the reflection amplitude at side
$L(R)$. The parameter $P_{n}$ corresponds to the spin-polarization of the
transmission probability $\left\vert t_{R(L),n\sigma}\right\vert ^{2}$. The
parameters $d\varphi_{n}^{L}$ and $d\varphi_{n}^{R}$ characterize the Spin
Dependence of Interfacial Phase Shifts (SDIPS), also called in other
references spin mixing angle\cite{Tokuyasu,Kopu,Zhao}. In our model, $P_{n}$
and $d\varphi_{n}^{L(R)}$ can be finite due to the spin-dependent interface
potential $\bar{V}_{b}$. Due to flux conservation and spin conservation along
$\vec{Z}$, the parameters $T_{n}$, $P_{n}$, $\varphi_{n}^{L(R)}$, and
$d\varphi_{n}^{L(R)}$ are sufficient to determine the value of the whole
$\mathcal{M}^{e}$ matrix (see Appendix \ref{Scattering} for details). Then,
using Eq.~(\ref{Mbar}), one can obtain an expression for $\bar{M}$. We will
work below at first order in $P_{n}$ and $d\varphi_{n}^{L(R)}$. In this case,
$\bar{M}$ can be decomposed as%
\begin{equation}
\bar{M}=\hat{M}^{0}(\mathbf{1}+\delta\bar{X}) \label{trail2}%
\end{equation}
The $n^{th}$ diagonal element of $\hat{M}^{0}$ in the transverse channel
subspace has the form, in the propagation direction subspace,
\begin{equation}
\hat{M}_{n,n}^{0}=\left[
\begin{array}
[c]{cc}%
\frac{ie^{i(\varphi_{n}^{L}+\varphi_{n}^{R})/2}}{\sqrt{T_{n}}} &
-ie^{i(\varphi_{n}^{R}-\varphi_{n}^{L})/2}\sqrt{\frac{R_{n}}{T_{n}}}\\
ie^{i(\varphi_{n}^{L}-\varphi_{n}^{R})/2}\sqrt{\frac{R_{n}}{T_{n}}} &
-\frac{ie^{-i(\varphi_{n}^{L}+\varphi_{n}^{R})/2}}{\sqrt{T_{n}}}%
\end{array}
\right]  \check{\sigma}_{0} \label{M0}%
\end{equation}
with $R_{n}=1-T_{n}$. Accordingly, the matrix $\delta\bar{X}$ is, in the
propagation direction subspace,
\begin{equation}
\delta\bar{X}=\left[
\begin{array}
[c]{cc}%
\delta\bar{X}_{++} & \delta\bar{X}_{+-}\\
\delta\bar{X}_{+-}^{\ast} & -\delta\bar{X}_{++}%
\end{array}
\right]  \label{dX1}%
\end{equation}
with
\begin{equation}
\delta\bar{X}_{n+,n+}=\frac{i\check{\sigma}_{Z}}{4T_{n}}\left(  T_{n}%
d\varphi_{n}^{L}+(2-T_{n})d\varphi_{n}^{R}\right)  \label{dX2}%
\end{equation}
and%
\begin{equation}
\delta\bar{X}_{n+,n-}=\frac{\check{\sigma}_{Z}e^{-i\varphi_{n}^{L}}}{2}\left(
\frac{P_{n}}{\sqrt{R_{n}}}-i\frac{\sqrt{R_{n}}}{T_{n}}d\varphi_{n}^{R}\right)
\,. \label{dX3}%
\end{equation}
One can check that Eqs.~(\ref{trail2})-(\ref{dX3}) are consistent with
Eq.~(\ref{FluxConservation}). Due to Eq.~(\ref{Mbar}), the matrices $\hat
{M}^{0}$ and $\delta\bar{X}$ are proportional to the identity in the Nambu
subspace. The matrix $\hat{M}^{0}$ is determined by the parameters $T_{n}$ and
$\varphi_{n}^{L(R)}$. It has a structure in the \underline{$\mathcal{E}$}
subspace only. In contrast, $\delta\bar{X}$ is a first order term in $P_{n}$
and $d\varphi_{n}^{L(R)}$, with a structure in the \underline{$\mathcal{E}$}
sub-space but also in the spin sub-space. We conclude that the matrices
$\hat{M}^{0}$ and $\check{G}_{L(R)}$ commute with each other, whereas
$\delta\bar{X}$ commutes neither with $\hat{M}^{0}$ nor with $\check{G}%
_{L(R)}$.

We want to express the matrix current of the isotropization zones as
\begin{equation}
\check{I}_{L(R)}(\varepsilon)=\check{I}_{L(R)}^{(0)}(\varepsilon)+\check
{I}_{L(R)}^{(1)}(\varepsilon) \label{matrixcurrentperturbation}%
\end{equation}
with $\check{I}_{L(R)}^{(0)}(\varepsilon)$ and $\check{I}_{L(R)}%
^{(1)}(\varepsilon)$ zeroth and first order terms in $\delta\bar{X}$,
respectively. We will mainly focus on the calculation of $\check{I}%
_{L}(\varepsilon)$ because the calculation of $\check{I}_{R}(\varepsilon)$ is
similar. To develop the expression (\ref{GBC}) of $\check{I}_{L}(\varepsilon
)$, one can use:
\begin{equation}
\widetilde{D}_{L}^{-1}=\widetilde{J}-\widetilde{J}\delta\widetilde
{V}\widetilde{J}+o(\delta\widetilde{V}^{2}) \label{den}%
\end{equation}
with%
\begin{equation}
\delta\widetilde{V}=\check{G}_{L}\check{G}_{R}\hat{Q}_{0}\text{ }\delta\bar
{X}+\check{G}_{L}\delta\bar{X}^{\dagger}\hat{Q}_{0}\check{G}_{R} \label{PERT4}%
\end{equation}%
\begin{equation}
\hat{Q}_{0}=(\hat{M}^{0})^{\dagger}\hat{M}^{0} \label{PERT1}%
\end{equation}
and%
\begin{equation}
\widetilde{J}=(\mathbf{1}+\hat{Q}_{0}\check{G}_{L}\check{G}_{R})^{-1}%
\end{equation}
For later use, we note that%
\begin{equation}
\widetilde{J}=\frac{\check{G}_{R}\check{G}_{L}+\hat{Q}_{0}^{-1}}{\left\{
\check{G}_{L},\check{G}_{R}\right\}  +\hat{Q}_{0}+\hat{Q}_{0}^{-1}}
\label{PERT2b}%
\end{equation}
In the next sections, we will substitute Eq.~(\ref{den}) into Eq.~(\ref{GBC}),
to express $\check{I}_{L}^{(0)}(\varepsilon)$ and $\check{I}_{L}%
^{(1)}(\varepsilon)$ in terms of the scattering parameters of the contact.

\subsection{Zeroth order component of the matrix current\label{WeakFerro0}}

We first discuss the conservation of the zeroth order matrix current across
the contact. From Eqs.~(\ref{rel}) and (\ref{Iiso}), one finds $\check{I}%
_{R}^{0}(\varepsilon)=2iG_{q}\mathrm{Tr}_{n,s}\left[  \hat{\Sigma}_{3}\hat
{M}^{0}\tilde{g}_{L}(\hat{M}^{0})^{\dag}\right]  /\pi$. Since $\hat{M}^{0}$
has a structure in the \underline{$\mathcal{E}$} subspace only, the cyclic
property of the trace $\mathrm{Tr}_{n,s}$ yields $\check{I}_{L}^{0}%
(\varepsilon)=\check{I}_{R}^{0}(\varepsilon)=\check{I}^{0}(\varepsilon)$.
Hence, the matrix current is conserved across the contact in the
spin-degenerate case.

We now calculate $\check{I}^{(0)}(\varepsilon)$. Since $\hat{M}^{0}$ commutes
with $\check{G}_{L(R)}$, Eq.~(\ref{GBC}) gives
\begin{equation}
\check{I}^{(0)}(\varepsilon)=2G_{q}\mathrm{Tr}_{n,s}\left[  \widetilde
{J}\left(  2\hat{\Sigma}_{3}\check{G}_{L}+\mathbf{1}-\hat{Q}_{0}\check{G}%
_{L}\check{G}_{R}\right)  \right]  . \label{Yulicurrent}%
\end{equation}
From Eq.(\ref{M0}), one finds
\begin{align}
\hat{Q}_{0}  &  =-2\hat{T}_{0}^{-1}\sqrt{\mathbf{1}-\hat{T}_{0}}\left[
\cos(\varphi_{n}^{L})\hat{\Sigma}_{1}+\sin(\varphi_{n}^{L})\hat{\Sigma}%
_{2}\right] \nonumber\\
&  +(2\hat{T}_{0}^{-1}-1)\hat{\Sigma}_{0} \label{Q0}%
\end{align}
and%
\begin{equation}
\hat{Q}_{0}^{-1}=\hat{\Sigma}_{3}\hat{Q}_{0}\hat{\Sigma}_{3} \label{Q0inv}%
\end{equation}
In Eq.~(\ref{Q0}), the matrices $\hat{\Sigma}_{0}$, $\hat{\Sigma}_{1}$ and
$\hat{\Sigma}_{2}$ refer to the identity, the first and second Pauli matrices
in the propagation direction subspace, respectively. We use $(\hat{T}%
_{0})_{ns,ms^{\prime}}^{\nu\sigma,\nu^{\prime}\sigma^{\prime}}=T_{n}%
\delta_{ss^{\prime}}\delta_{nm}\delta_{\sigma\sigma^{\prime}}\delta_{\nu
\nu^{\prime}}\mathbf{1}_{K}$. We find that $\hat{Q}_{0}+\hat{Q}_{0}^{-1}$ has
a diagonal structure in the propagation direction space. Therefore, using
expression (\ref{PERT2b}) for $\widetilde{J}$, and performing the trace over
the channel and propagation direction indices, we obtain
\begin{equation}
\check{I}^{(0)}(\varepsilon)=4G_{q}%
{\textstyle\sum\limits_{n}}
\frac{T_{n}\left[  \check{G}_{R},\check{G}_{L}\right]  }{4+T_{n}(\left\{
\check{G}_{L},\check{G}_{R}\right\}  -2)} \label{Yulicurrent_comutator}%
\end{equation}
Equation~(\ref{Yulicurrent_comutator}) corresponds to the expression obtained
in Ref.~\onlinecite{RefYuli} for a spin-independent contact\cite{YipComment}.
This expression does not involve any scattering phase shift.

\subsection{First order component of the matrix current\label{WeakFerro1}}

We now concentrate on the contribution $\check{I}_{L(R)}^{(1)}(\varepsilon)$
to the matrix current to first order in $\delta\bar{X}$. Equations (\ref{GBC})
and (\ref{den}) lead to
\begin{equation}
\check{I}_{L}^{(1)}(\varepsilon)=-4G_{q}\mathrm{Tr}_{n,s}\left[  \widetilde
{J}\text{ }\delta\widetilde{V}\text{ }\widetilde{J}\left(  \mathbf{1}%
+\hat{\Sigma}_{3}\text{ }\check{G}_{L}\right)  \right]
\label{matrixcurrentperturbation1}%
\end{equation}
with $\delta\widetilde{V}$ given by Eq.~(\ref{PERT4}). Using Eqs.~(\ref{dX1}%
-\ref{dX3}) and (\ref{Q0}-\ref{Q0inv}), and performing the trace over the
transverse channel and propagation direction indices (see Appendix
\ref{MatrixCurrent} for details), one finds
\begin{align}
\check{I}_{L}^{(1)}(\varepsilon)  &  =2G_{q}\sum_{n}\left(  4+T_{n}(\left\{
\check{G}_{L},\check{G}_{R}\right\}  -2)\right)  ^{-1}\nonumber\\
&  \times\left(  4T_{n}P_{n}\left[  \left\{  \check{\sigma}_{Z},\check{G}%
_{R}\right\}  ,\check{G}_{L}\right]  -i8R_{n}d\varphi_{n}^{L}\left[
\check{\sigma}_{Z},\check{G}_{L}\right]  \right. \nonumber\\
&  +iT_{n}\left(  T_{n}d\varphi_{n}^{L}+(2-T_{n})d\varphi_{n}^{R}\right)
\left[  \check{G}_{R}\left[  \check{\sigma}_{Z},\check{G}_{R}\right]
,\check{G}_{L}\right] \nonumber\\
&  \left.  -iT_{n}\left(  T_{n}d\varphi_{n}^{R}+\left(  2-T_{n}\right)
d\varphi_{n}^{L}\right)  \left[  \left[  \check{\sigma}_{Z},\check{G}%
_{R}\right]  \check{G}_{L},\check{G}_{L}\right]  \right) \nonumber\\
&  \times\left(  4+T_{n}(\left\{  \check{G}_{L},\check{G}_{R}\right\}
-2)\right)  ^{-1} \label{I_1_final}%
\end{align}
A comparison between Eqs.~(\ref{GBC}) and (\ref{GBC3}) indicates that the
expression of $\check{I}_{R}^{(1)}(\varepsilon)$ can be obtained by
multiplying the expression (\ref{I_1_final}) of $\check{I}_{L}^{(1)}%
(\varepsilon)$ by $-1$, replacing $d\varphi_{n}^{L(R)}$ by $d\varphi
_{n}^{R(L)}$, and $\check{G}_{L(R)}$ by $\check{G}_{R(L)}$. Note that the
expressions of $\check{I}_{L}^{(1)}(\varepsilon)$ and $\check{I}_{R}%
^{(1)}(\varepsilon)$ involve the SDIPS parameters $d\varphi_{n}^{L}$ and
$d\varphi_{n}^{R}$ but not the spin-averaged phases $\varphi_{n}^{L}$ and
$\varphi_{n}^{R}$.

\subsection{Expression of the matrix current in the tunnel
limit\label{weakferrotunnel}}

We now assume that the contact is a tunnel barrier $(T_{n}\ll1)$, which seems
reasonable considering the band structure mismatch between most $S$ and $F$
materials. At first order in $T_{n}$, the matrix currents $\check{I}%
_{L(R)}(\varepsilon)$ take the very transparent form%
\begin{align}
2\check{I}_{L}(\varepsilon)  &  =G_{T\text{ }}\left[  \check{G}_{R},\check
{G}_{L}\right]  +G_{MR\text{ }}\left[  \left\{  \check{\sigma}_{Z},\check
{G}_{R}\right\}  ,\check{G}_{L}\right] \nonumber\\
&  +iG_{\phi}^{L}\left[  \check{\sigma}_{Z},\check{G}_{L}\right]  +iG_{\chi
}^{L}[\check{G}_{R}\check{G}_{L}\check{\sigma}_{Z}+\check{\sigma}_{Z}\check
{G}_{L}\check{G}_{R},\check{G}_{L}]\nonumber\\
&  +iG_{\chi}^{R}\left[  \check{G}_{R}\left[  \check{\sigma}_{Z},\check{G}%
_{R}\right]  ,\check{G}_{L}\right]  \label{Iprl}%
\end{align}
and
\begin{align}
2\check{I}_{R}(\varepsilon)  &  =G_{T\text{ }}\left[  \check{G}_{R},\check
{G}_{L}\right]  +G_{MR\text{ }}\left[  \check{G}_{R},\left\{  \check{\sigma
}_{Z},\check{G}_{L}\right\}  \right] \nonumber\\
&  -iG_{\phi}^{R}\left[  \check{\sigma}_{Z},\check{G}_{R}\right]  -iG_{\chi
}^{R}[\check{G}_{L}\check{G}_{R}\check{\sigma}_{Z}+\check{\sigma}_{Z}\check
{G}_{R}\check{G}_{L},\check{G}_{R}]\nonumber\\
&  -iG_{\chi}^{L}\left[  \check{G}_{L}\left[  \check{\sigma}_{Z},\check{G}%
_{L}\right]  ,\check{G}_{R}\right]  \,. \label{Iprr}%
\end{align}
We have introduced above the conductance parameters\cite{Erratum}
\begin{align}
G_{T}/G_{q}  &  =2\sum\nolimits_{n}T_{n}\label{GT}\\
G_{MR\text{ }}/G_{q}  &  =\sum\nolimits_{n}T_{n}P_{n}\label{GMR}\\
G_{\phi}^{L(R)}/G_{q}  &  =-2\sum\nolimits_{n}d\varphi_{n}^{L(R)}\label{Gfi}\\
G_{\chi}^{L(R)}/G_{q}  &  =\sum\nolimits_{n}T_{n}d\varphi_{n}^{L(R)}/2
\label{Gchi}%
\end{align}
The values of the coefficients $G_{T}$, $G_{MR}$, $G_{\phi}^{L(R)}$, and
$G_{\chi}^{L(R)}$ are difficult to predict because they depend on the detailed
microscopic structure of the interface. These parameters can in principle be
large compared to $G_{q}$ because, although the derivation of Eqs.~(\ref{Iprl}%
) and (\ref{Iprr}) assumes that $T_{n}$, $P_{n}$ and $d\varphi_{n}^{L(R)}$ are
small, the definitions (\ref{GT})-(\ref{Gchi}) involve a summation on a
numerous number of channels. The parameter $G_{MR\text{ }}$ can be finite when
$P_{n}\neq0$ and the parameters $G_{\phi}^{L(R)}$ and $G_{\chi}^{L(R)}$ can be
finite due to the SDIPS. From Eqs.~(\ref{GT})-(\ref{Gchi}), $G_{\chi}^{L}$ and
$G_{\chi}^{R}$ are likely to be small compared to $G_{T}$ and $G_{\phi}%
^{L(R)}$. This is why these coefficients were disregarded so far for studying
the effects of the SDIPS on the superconducting proximity effect. In contrast,
it is possible to have $G_{\phi}^{L(R)}>G_{T}$ as well as $G_{\phi}%
^{L(R)}<G_{T}$, using a spin-dependent interface potential $\bar{V}_{b}%
$\cite{Cottet2}. We also note that the hypothesis $P_{n}\ll1$ imposes
$G_{MR\text{ }}\ll G_{T}$. We have checked that in the normal-state limit,
Eqs.~(\ref{Iprl},\ref{Iprr}) agree with the boundary conditions introduced in
Refs. \onlinecite{RefCircuit} and \onlinecite{Braatas} provided the reflection
and transmission mixing conductances $G_{mix}^{L(R),r}$ and $G_{mix}^{t}$
appearing in these boundary conditions are replaced by their developments at
first order in $T_{n}$, $P_{n}$, and $d\varphi_{n}^{L(R)}$ i.e.:%
\[
G_{mix}^{L(R),r}\rightarrow(G_{T}/2)+i(G_{\phi}^{L(R)}/2)+2iG_{\chi}^{L(R)}%
\]
and
\[
G_{mix}^{t}\rightarrow(G_{T}/2)+i(G_{\chi}^{L}+G_{\chi}^{R})
\]
(see Appendix \ref{NormalState} for details).

We now briefly review the physical effects of the coefficients $G_{T}$,
$G_{MR}$, and $G_{\phi}^{L(R)}$. The term in $G_{T}$ in Eqs.~(\ref{Iprl}) and
(\ref{Iprr}) corresponds to the term derived in Ref.~\onlinecite{Kuprianov}
for superconducting/normal metal interfaces. This term is responsible for the
superconducting proximity effect occurring in a normal metal layer or a
ferromagnetic layer in contact with a superconductor. The parameter $G_{MR}$
accounts for the spin-dependence of the contact tunnel probabilities, and thus
leads to magnetoresistance effects\cite{Morten1,Morten2,DiLorenzo}. In a
ferromagnet $F$ subject to the proximity effect, the ferromagnetic exchange
field causes spatial oscillations of the isotropic Green's function $\check
{G}$, which results e.g. in spatial oscillations of the density of states of
$F$. It has been shown that the $G_{\phi}^{L(R)}$ terms can shift these
oscillations\cite{Cottet2,Cottet1,Cottet3}. The $G_{\phi}^{L(R)}$ terms also
induce Zeeman effective fields inside thin superconducting or normal metal
layers\cite{Dani1,Dani2,Cottet2}. In principle, in non-collinear geometries
enclosing several ferromagnetic elements with non-collinear magnetizations,
the SDIPS terms can induce spin-precession effects.

Note that, so far, we have considered that the interface potential $\bar
{V}_{b}$ is spin-polarized along the $\vec{Z}$ direction. In the general case,
due to interface effects, it is possible that the spin-dependent part of the
interface potential $\bar{V}_{b}$ is polarized along a direction $\vec{m}$
different from the bulk exchange field direction of contacts $L$ or $R$. It is
also possible that the contact belongs to a circuit enclosing several
ferromagnets with non-collinear magnetizations, or ferromagnets with a
spatially dependent magnetization direction. In these cases, Eqs.~(\ref{GBC}%
-\ref{der}) are still valid. One can furthermore generalize the BCIGF
(\ref{I_1_final}), (\ref{Iprl}) and (\ref{Iprr}) to an arbitrary spin
reference frame ($\check{\sigma}_{X},\check{\sigma}_{Y},\check{\sigma}_{Z}$)
by replacing $\check{\sigma}_{Z}$ by $\left(  (1+\check{\tau}_{3})\sigma
_{Z}(\vec{m}\cdot\check{\vec{\sigma}})\sigma_{Z}+(\check{\tau}_{3}%
-1)\sigma_{y}(\vec{m}\cdot\check{\vec{\sigma}})\sigma_{y}\right)  /2$.

As we have already explained in Sec. \ref{ballistic}, the use of transfer
matrices for the derivation of Eqs.~(\ref{Iprl}-\ref{Gchi}) allows to obtain
results for the $T_{n}\rightarrow0$ limit, which must be performed after an
explicit calculation of the BCIGF. From Eq.~(\ref{Gfi}), even if a channel $n$
is perfectly reflected at the $L/R$ boundary, it can contribute to the matrix
current due to the spin dependence of the reflection phase $d\varphi
_{n}^{L(R)}$. We will recover this result in Sec. \ref{SFIbcs} for a $S/FI$
contact, using an approach without transfer matrices.

\subsection{Discussion on the matrix current conservation and the
spin-dependent circuit theory\label{chargeCurrent}}

In this section, we discuss the non-conservation of the matrix current in the
general case. We have already seen in section \ref{WeakFerro0} that the full
matrix current is conserved across an interface in the spin degenerate case.
In the spin-dependent situation, one finds from Eqs.~(\ref{rel}) and
(\ref{Iiso}) that $\check{I}_{R}(\varepsilon)=2iG_{q}\mathrm{Tr}_{n,s}\left[
\hat{\Sigma}_{3}\bar{M}\tilde{g}_{L}(\bar{M})^{\dag}\right]  /\pi$. Since
$\bar{M}$ has a structure in the spin subspace, the cyclic property of the
trace $\mathrm{Tr}_{n,s}$ cannot be used anymore to relate $\check{I}%
_{L}(\varepsilon)$ and $\check{I}_{R}(\varepsilon)$. Hence, nothing imposes
$\check{I}_{L}(\varepsilon)=\check{I}_{R}(\varepsilon)$ in the general case.
Reference \onlinecite {Cottet2} illustrates that in the case of a simple $S/F$
bilayer with a homogeneous magnetization in $F$, $\check{I}_{L}(\varepsilon
)\neq\check{I}_{R}(\varepsilon)$ is already possible. Note that $\check{I}%
_{L}(\varepsilon)\neq\check{I}_{R}(\varepsilon)$ does not violate particle
current conservation through the interface, although the average current
flowing at side $Q$ of the contact is determined by $\check{I}_{Q}$, i.e.
\begin{equation}
\left\langle I_{Q}\right\rangle =\frac{1}{16\mathrm{e}}\int_{-\infty}^{\infty
}d\varepsilon\mathrm{Tr}_{\nu\sigma}\left\{  \check{\tau}_{3}\check{I}_{Q}%
^{K}(\varepsilon)\right\}  \label{current}%
\end{equation}
Indeed, the above equation leads to
\[
\left\langle I_{L}\right\rangle =\frac{G_{q}}{8\mathrm{e}}\int_{-\infty
}^{\infty}d\varepsilon\mathrm{Tr}_{ns\nu\sigma}\left\{  \check{\tau}_{3}%
\hat{\Sigma}_{3}\tilde{g}_{L}^{K}\right\}
\]
and%
\[
\left\langle I_{R}\right\rangle =\frac{G_{q}}{8\mathrm{e}}\int_{-\infty
}^{\infty}d\varepsilon\mathrm{Tr}_{ns\nu\sigma}\left\{  \check{\tau}_{3}%
\hat{\Sigma}_{3}\bar{M}\tilde{g}_{L}^{K}\bar{M}^{\dag}\right\}
\]
Since $\bar{M}$ is proportional to the identity in the Keldysh space, one can
use the cyclic property of the trace $Tr_{ns\nu\sigma}$ in the above
equations, to show that $\left\langle I_{L}\right\rangle =\left\langle
I_{R}\right\rangle $. It is important to point out that \textit{the
non-conservation of the matrix current at the }$L/R$\textit{ boundary does not
affect the applicability of Eqs. }(\ref{GBC}-\ref{der}). The fact that the
matrix current is not conserved through a spin-dependent interface has the
obvious reason that only charge conservation is required by fundamental laws,
whereas other quantities are not conserved in general. It depends on the
symmetry of the Hamiltonian describing the barrier, which quantities are
conserved in addition to charge. If the barrier potential is spin-independent,
all elements of the matrix current are conserved. In general, this is not the
case anymore for spin-dependent barriers. An extreme case illustrating this
situation is provided by an interface between a $FI$ and a metal. In the $FI$,
the concept of a matrix current does not even exist, although the $FI$
influences the adjacent metal due to the proximity effect. We will discuss
this case in section \ref{SFI}.

The BCIGF derived in this article allow to generalize the "circuit theory" of
Ref. \onlinecite{RefYuli} to the case of multiterminal circuits which enclose
superconductors, normal metals, ferromagnets and ferromagnetic insulators. In
the approach of circuit theory, a system is split up into reservoirs $r$
(voltage sources), connectors $c$ (contacts, interfaces) and nodes $n$ (small
islands) in analogy to classical electric circuits. Each reservoir or node is
characterized with an isotropic Green's function with no space dependence,
which plays the role of a generalized potential. Circuit theory requires to
apply generalized Kirchhoff's rules on the matrix current $\check{I}$. We have
seen above that $\check{I}$ is not conserved through the contacts in the
general case, but this is not a problem since we know how to express the
matrix current at both sides of the contact. We will note $\check{I}_{c}^{n}$
the matrix current flowing from the connector $c$ into node $n$, which is
given by Eqs.(\ref{GBC}) or (\ref{GBC3}). One must be careful to the fact that
the matrix current is not conserved either inside the nodes due the terms on
the right hand side of the Usadel Eq.~(\ref{Usadel1}). To compensate for the
non-conservation of $\check{I}$ inside node $n$, one can introduce a leakage
matrix current
\begin{equation}
\check{I}_{leakage}^{n}=4\pi G_{q}\nu_{0}V_{n}\left[  -i\varepsilon\check
{\tau}_{3}+\check{\Delta}+iE_{ex}\check{\sigma}_{Z},\check{G}_{n}\right]
\end{equation}
which accounts for the \textquotedblleft leakage\textquotedblright\ of
quantities like for instance electron-hole coherence or spin accumulation. In
the above expression, $\check{G}_{n}$, $\check{\Delta}$, and $E_{ex}$, refer
to the values of the isotropic Green's function, gap matrix, and exchange
field inside $n$, and $V_{n}$ is the volume of the node. The leakage matrix
current $\check{I}_{leakage}^{n}$ can be viewed as flowing from an effective
"leakage terminal". It must occur in the generalized Kirchhoff's rule for node
$n$, i.e.
\[
\check{I}_{leakage}^{n}+\sum\nolimits_{c}\check{I}_{c}^{n}=0
\]
with the index $c$ running on all the contacts connected to node $n$. We refer
the reader to Refs. \onlinecite{RefYuli,YuliBook} for more details on circuit theory.

\section{Contact between a metal and a ferromagnetic insulator\label{SFI}}

\subsection{Boundary conditions\label{SFIbcs}}

In the case of a contact between a metal and a ferromagnetic insulator, one
can perform a calculation similar to the one of the metallic case without
using the transfer matrix $\bar{M}$ but a simpler "pseudo" transfer matrix
$\mathbb{\bar{M}}$ which involves only reflexion phases against the $FI$ (see
definition below). This facilitates a developement of the BCIGF at higher
orders in the SDIPS parameters. We assume that the ferromagnetic insulator is
located at the right side ($z>0$) of the contact, and that the left side $L$
is a BCS superconductor, a normal metal, or a ferromagnet, which satisfies Eq.
(\ref{QC}). We divide $L$ into a ballistic zone, an isotropization zone and a
diffusive zone like in Figure~\ref{Dessin}. We directly consider the case
where the contact locally conserves the transverse channel index and spins
along $\vec{Z}$. In this case, one can introduce a reflection phase shift
$\varphi_{n}+\sigma d\varphi_{n}/2$ such that the left-going and right-going
quasiparticle wavefunctions in the $n^{\text{th}}$ channel of $L$ are related
by
\[
\psi_{n,-}^{\nu,\sigma}(-c_{L},\varepsilon)=e^{i(\varphi_{n}+\sigma
\frac{d\varphi_{n}}{2})}\psi_{n,s+}^{\nu,\sigma}(-c_{L},\varepsilon)
\]
Using this relation, one can check that the calculations of
sections~\ref{ballistic} to \ref{generalBC} can be repeated by replacing the
ballistic Green's function $\tilde{g}_{R}$ by $\hat{\Sigma}_{1}\tilde{g}%
_{L}\hat{\Sigma}_{1}$, $\check{G}_{R}$ by $\check{G}_{L}$, and the transfer
matrix $\bar{M}$ by a pseudo transfer matrix $\mathbb{\bar{M}}=\mathbb{\hat
{M}}^{0}\exp(\delta\mathbb{\bar{X}})$. The $n^{th}$ diagonal elements of
$\mathbb{\hat{M}}^{0}$ and $\delta\mathbb{\bar{X}}$ in the transverse channel
subspace write%
\begin{equation}
\mathbb{\hat{M}}_{n}^{0}=\left[  \cos\left(  \varphi_{n}\right)  \hat{\Sigma
}_{0}+i\sin\left(  \varphi_{n}\right)  \hat{\Sigma}_{3}\right]  \check{\sigma
}_{0}%
\end{equation}
and
\begin{equation}
\delta\mathbb{\bar{X}}_{n}=id\varphi_{n}\hat{\Sigma}_{3}\check{\sigma}%
_{Z}/2\,.
\end{equation}
Since $\mathbb{\hat{M}}^{0}$ commutes with $\check{G}_{L}$ and $\delta
\mathbb{\bar{X}}$ and $(\mathbb{\hat{M}}^{0})^{\dag}\mathbb{\hat{M}}%
^{0}=\mathbf{1}$, we find%
\begin{equation}
\check{I}_{L}(\varepsilon)=2G_{q}\mathrm{Tr}_{n,s}\left[  \left(
\mathbf{1}+\delta\widetilde{Y}\right)  ^{-1}\left(  \check{G}_{L}\hat{\Sigma
}_{3}+\mathbf{1}\right)  -\mathbf{1}\right]  \label{IfiGal}%
\end{equation}
with%
\begin{equation}
\delta\widetilde{Y}\mathbb{=(}\check{G}_{L}e^{-\delta\mathbb{\bar{X}}}%
\check{G}_{L}e^{\delta\mathbb{\bar{X}}}-\mathbf{1})/2 \label{IfiGal2}%
\end{equation}
Hence, quite generally, the spin-averaged reflection phases $\varphi_{n}$ do
not contribute to $\check{I}_{L}(\varepsilon)$. Equation~(\ref{IfiGal}) can be
traced out numerically. Alternatively, one can achieve further analytical
progress by expanding $\check{I}_{L}(\varepsilon)$ with respect to the
spin-dependent part $\delta\widetilde{Y}$. We have $\left(  \mathbf{1}%
+\delta\widetilde{Y}\right)  ^{-1}=\mathbf{1}+\sum\nolimits_{n}\left(
-\delta\widetilde{Y}\right)  ^{n}$. Therefore, at fourth order in
$d\varphi_{n}$ we obtain
\begin{align}
2\check{I}_{L}(\varepsilon)  &  =iG_{\phi,1}\left[  \check{\sigma}_{Z}%
,\check{G}_{L}\right]  +G_{\phi,2}\left[  \check{\sigma}_{Z},\check{G}%
_{L}\check{\sigma}_{Z}\check{G}_{L}\right] \nonumber\\
&  +iG_{\phi,3}\left[  \check{\sigma}_{Z},\check{G}_{L}\left(  \check{\sigma
}_{Z}\check{G}_{L}\right)  ^{2}\right] \nonumber\\
&  +G_{\phi,4}\left[  \check{\sigma}_{Z},\check{G}_{L}\left(  \check{\sigma
}_{Z}\check{G}_{L}\right)  ^{3}\right]  \label{bcSFI}%
\end{align}
with the conductance parameters
\begin{align}
G_{\phi,1}/G_{q}  &  =-2\sum\nolimits_{n}d\varphi_{n}-\sum\nolimits_{n}%
d\varphi_{n}^{3}/24\\
G_{\phi,2}/G_{q}  &  =\sum\nolimits_{n}d\varphi_{n}^{2}/2+\sum\nolimits_{n}%
d\varphi_{n}^{4}/48\\
G_{\phi,3}/G_{q}  &  =\sum\nolimits_{n}d\varphi_{n}^{3}/8\\
G_{\phi,4}/G_{q}  &  =-\sum\nolimits_{n}d\varphi_{n}^{4}/32\,.
\end{align}
In the normal-state limit, we have checked that Eq.~(\ref{bcSFI}) agrees with
the BCIGF presented in Refs.~\onlinecite{RefCircuit} and \onlinecite{Braatas}
(see appendix \ref{NormalState} for a detailed comparison). The term in
$G_{\phi,1}$ of Eq.~(\ref{bcSFI}) has already been used in
Refs.~\onlinecite {Dani1} and \onlinecite{Dani2}. At first order in
$d\varphi_{n}$, it is the only term contributing to $\check{I}_{L}$, and it
can be recovered from Eqs.~(\ref{Iprl}) and (\ref{GT})-(\ref{Gchi}) by using
$T_{n}=0$ and $d\varphi_{n}^{L}=d\varphi_{n}$. At higher orders in
$d\varphi_{n}$, the value of $G_{\phi,1}$ is renormalized and new terms occur
in the expression of $\check{I}_{L}$. The second order term has a
straightforward interpretation, since it has exactly the same matrix structure
as the self-energy due to scattering by paramagnetic impurities in a normal
metal\cite{Maki,AG}, or due to magnetic disorder along the $\overrightarrow
{Z}$ direction in a ferromagnet\cite{Houzet}. The scattering of Cooper pairs
at the spin-active interface leads to a coupling between spin-singlet and
spin-triplet components, which, due to the random scattering at second order
leads to pair breaking. In a similar fashion, we can understand the higher
order terms in Eq.~\ref{bcSFI} as a result of multiple scattering at the
$S/FI$ interface. Note that in this section, we have assumed that the $FI$
side of the contact is magnetized along the $\vec{Z}$ direction. If the $FI$
is magnetized along a direction $\vec{m}\neq\vec{Z}$, one can describe the
contact in the spin reference frame $(\check{\sigma}_{X},\check{\sigma}%
_{Y},\check{\sigma}_{Z})$ by replacing $\check{\sigma}_{Z}$ by $\left(
(1+\check{\tau}_{3})\sigma_{Z}(\vec{m}\cdot\check{\vec{\sigma}})\sigma
_{Z}+(\check{\tau}_{3}-1)\sigma_{y}(\vec{m}\cdot\check{\vec{\sigma}}%
)\sigma_{y}\right)  /2$ in Eq. (\ref{bcSFI}).

\subsection{Example of a $S/FI$ bilayer}

To illustrate some effects of the $G_{\phi,i}$ coefficients, we now consider
the case of a $S/FI$ bilayer, with $S$ located at $z\in\lbrack0,d_{S}]$ and
$FI$ at $z>d_{S}$. Throughout this section, we replace the energy
$-i\varepsilon$ appearing in the Usadel equation by $-i\varepsilon+\Gamma$,
where the phenomenological collision rate $\Gamma$ accounts for inelastic
processes \cite{thesisWB}. Inside $S$, the retarded part of the isotropic
Green's function can be parametrized with a so-called pairing angle
$\Lambda_{\sigma}^{S}$ such that
\[
\check{G}^{r}=\left[
\begin{array}
[c]{cccc}%
\cos(\Lambda_{\uparrow}^{S}) & 0 & 0 & \sin(\Lambda_{\uparrow}^{S})\\
0 & \cos(\Lambda_{\downarrow}^{S}) & \sin(\Lambda_{\downarrow}^{S}) & 0\\
0 & \sin(\Lambda_{\downarrow}^{S}) & -\cos(\Lambda_{\downarrow}^{S}) & 0\\
\sin(\Lambda_{\uparrow}^{S}) & 0 & 0 & -\cos(\Lambda_{\uparrow}^{S})
\end{array}
\right]
\]
Let us first assume that $d_{S}\ll\xi_{S}$, so that one can use the quadratic
approximation $\Lambda_{\sigma}^{S}(\varepsilon,x)=\Lambda_{\sigma}^{0}%
-\beta_{\sigma}(x/\xi_{S})^{2}$ and a constant superconducting gap
$\Delta(x)=\Delta_{0}$ inside $S$ (see e.g. Ref.~\onlinecite {Cottet2}). For
$z\in\lbrack0,d_{S}]$, the Usadel equations (see appendix \ref{UsadelAppendix}%
) lead to
\begin{equation}
\beta_{\sigma}=\frac{\Delta_{0}\cos(\Lambda_{\sigma}^{0})+\left(
i\varepsilon-\Gamma\right)  \sin(\Lambda_{\sigma}^{0})}{2\Delta_{BCS}}
\label{Eq1}%
\end{equation}
We have introduced above the bulk BCS gap $\Delta_{BCS}$ of $S$. The value of
$\Lambda_{\sigma}^{0}$ can be found by identifying Eq.~(\ref{Eq1}) with
Eq.~(\ref{bcSFI}), i.e.
\begin{align}
2\beta_{\sigma}d_{s}/\xi_{S}  &  =i\gamma_{\phi,1}\sigma\sin(\Lambda_{\sigma
}^{0})+\gamma_{\phi,2}\sin(2\Lambda_{\sigma}^{0})\nonumber\\
&  +i\gamma_{\phi,3}\sigma\sin(3\Lambda_{\sigma}^{0})+\gamma_{\phi,4}%
\sin(4\Lambda_{\sigma}^{0}) \label{Eq2}%
\end{align}
(see Appendix \ref{GorkovToUsadel} for details). We have introduced above
$\gamma_{\phi,i}=G_{\phi,i}\xi_{S}\rho_{S}/A$. Note that the value of
$\Delta_{0}$ must be calculated self-consistently with $\Lambda_{\sigma}^{0}$,
see e.g. Ref.~\onlinecite {Cottet2}. We will first consider the case
$G_{\phi,2}=G_{\phi,3}=G_{\phi,4}=0$, for which Eqs.~(\ref{Eq1}-\ref{Eq2})
yield
\begin{equation}
\Lambda_{\sigma}^{0}=\arctan\left(  \frac{\Delta_{0}}{-i\varepsilon
+\Gamma+i\gamma_{\phi,1}\sigma\frac{\xi_{S}}{d_{s}}\Delta_{BCS}}\right)
\label{split}%
\end{equation}
From the above Eq., $G_{\phi,1}$ induces an effective Zeeman field
$H_{eff}=2i\gamma_{\phi,1}\xi_{S}\Delta_{BCS}/d_{s}g\mu_{B}$ inside a thin $S$
layer, like the $G_{\phi}^{L(R)}$ terms of section \ref{weakferrotunnel}
\cite{Dani1,Dani2,Cottet2}. \begin{figure}[ptbh]
\includegraphics[width=0.9\linewidth]{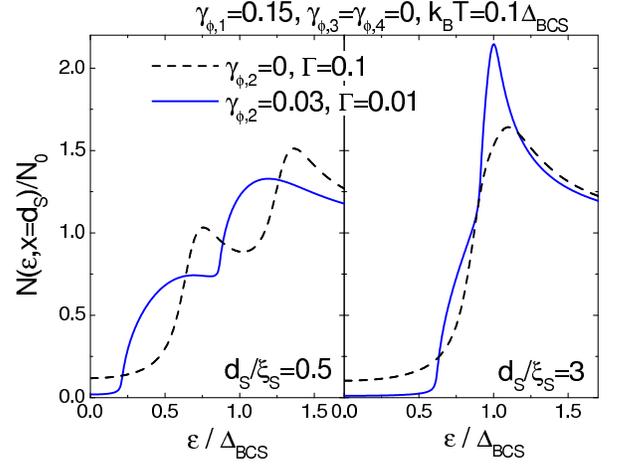}\caption{Density of states
$N(\varepsilon,x=d_{S})$ in a $S$ layer contacted to a $FI$. The black dashed
lines correspond to $\Gamma=0.1$ and $\gamma_{\phi,2}=0$, and the blue full
lines correspond to $\Gamma=0.01$ and $\gamma_{\phi,2}=0.03$. The left panel
corresponds to $d_{S}/\xi_{S}=0.5$ and the right panel corresponds to
$d_{S}/\xi_{S}=3$. In all cases, we have used $\gamma_{\phi,1}=0.15$,
$\gamma_{\phi,3}=\gamma_{\phi,4}=0$ and $k_{B}T=0.1\Delta_{BCS}$.}%
\label{NUM1}%
\end{figure}The density of states (DOS) in the $S$ layer can be calculated as
$N(\varepsilon,x)=N_{0}\sum\nolimits_{\sigma}\operatorname{Re}[\cos
(\Lambda_{\sigma}^{S}(\varepsilon,x))]/2$, with $N_{0}$ the normal-state
density of states. The $G_{\phi,1}$-induced effective Zeeman field $H_{eff}$
splits the superconducting peaks of the DOS, as shown by the black dashed line
in Fig. \ref{NUM1}, left panel. Spin-splitting effects in $S/F$ systems were
first intuited by De Gennes from a generalization of Cooper's
argument\cite{Cooper,DeGennes}. Later, Ref. \onlinecite{Tokuyasu} has
confirmed from a quasiclassical approach that the SDIPS can induce a
spin-splitting of the DOS in a ballistic $S/FI$ bilayer with a thin $S$.
However, the effect found by Tokuyasu \textit{et al.} is qualitatively
different from ours. Indeed, in the ballistic limit, Tokuyasu \textit{et al.}
find that the $S/FI$ bilayer differs from a $S$ layer in an external field
because the SDIPS induced spin-splitting effect depends upon the quasiparticle
trajectory. In contrast, in the diffusive limit, we obtain a true effective
Zeeman field $H_{eff}$ which appears directly in the spectral functions. On
the experimental side, spin-splitted DOS were observed in superconducting Al
layers contacted to different types of $FI$ as soon as 1986 (see
Refs.~\onlinecite{Tedrow,Meservey,MooderaTheoDiff,Hao}). However, the
inadequacy of the ballistic approach of Tokuyasu et al. for modeling the
actual experiments was pointed out in Ref. \onlinecite{Hao}. In fact, most of
the experiments on Al$/FI$ interfaces were interpreted by their authors in
terms of a diffusive approach with no SDIPS, and an internal Zeeman field
added arbitrarily in the Al layer (see Refs.
\onlinecite{MooderaTheoDiff,Hao,Alexander}). Our approach provides a
microscopic justification for the use of such an internal field. Remarkably,
it was found experimentally\cite{Hao} that the internal field appearing in
$S$\ scales with $d_{s}^{-1}$, in agreement with our expression of $H_{eff}%
$.\begin{figure}[ptbh]
\includegraphics[width=0.7\linewidth]{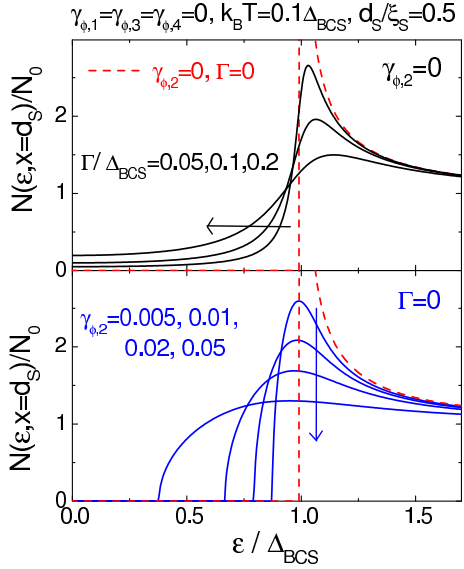}\caption{Top panel: Density of
states $N(\varepsilon,x=d_{S})$ in a $S$ layer contacted to a $FI$. The red
dashed line corresponds to $\Gamma=0$ and $\gamma_{\phi,2}=0$ in both panels.
The black full lines in the top panel correspond to $\Gamma\neq0$ and
$\gamma_{\phi,2}=0$ and the blue full lines in the bottom panel correspond to
$\Gamma=0$ and $\gamma_{\phi,2}\neq0$. In all cases, we have used $d_{S}%
/\xi_{S}=0.5$, $\gamma_{\phi,1}=\gamma_{\phi,3}=\gamma_{\phi,4}=0$ and
$k_{B}T=0.1\Delta_{BCS}$.}%
\label{NUM2}%
\end{figure}

We now discuss briefly the effects of the $G_{\phi,2}$, $G_{\phi,3}$ and
$G_{\phi,4}$ terms. Assuming $\Lambda_{\sigma}^{0}\ll2\pi$, the linearization
of Eq. (\ref{Eq2}) leads to
\begin{equation}
\Lambda_{\sigma}^{0}=\frac{\Delta_{0}}{-i\varepsilon+\Gamma+\xi_{S}%
\Delta_{BCS}\frac{i\left(  \gamma_{\phi,1}+3\gamma_{\phi,3}\right)
\sigma+2\gamma_{\phi,2}+4\gamma_{\phi,4}}{d_{s}}}\,.
\end{equation}
Therefore, in this limit, $G_{\phi,3}$ contributes to the Zeeman effective
field like $G_{\phi,1}$. Moreover, the coefficients $G_{\phi,2}$ and
$G_{\phi,4}$ lead to a decoherence effect similar to the decoherence induced
by the $\Gamma$ term. However, it is clear from Eq. (\ref{Eq2}) that this
picture is not valid in the general case. Let us focus on the effect of
$G_{\phi,2}$. From (\ref{Eq2}), in the non-linearized limit, $\gamma_{\phi,2}$
occurs together with a $\sin(2\Lambda_{\sigma}^{0})$ in the expression of
$\beta_{\sigma}$. Therefore, as already pointed out in section \ref{SFIbcs},
in the general case, it is more relevant to compare the effect of
$\gamma_{\phi,2}$ to that of paramagnetic impurities which would be diluted
inside $S$. The analogy to magnetic disorder can be understood as arising due
to successive reflections on the $S/FI$ interface with random spin-dependent
phase shifts. To study the effect of $\gamma_{\phi,2}$ in the general case, we
have calculated the density of states $N(\varepsilon,x)$ numerically. Our code
takes into account the self-consistency of the superconducting gap $\Delta(x)$
in the $S$ layer and is valid for arbitrary values of $d_{S}$\cite{code}.
Figure \ref{NUM2} compares the effect of $\Gamma\neq0$ (top panel) with the
effect of $G_{\phi,2}\neq0$ (bottom panel), for $G_{\phi,1}=0$. As expected,
we find that the effect of $G_{\phi,2}$ on the DOS of a thin $S$ is quite
similar to the effect of paramagnetic impurities which would be diluted inside
the bulk of $S$\cite{Skalski,Ambegaokar}. \begin{figure}[ptbh]
\includegraphics[width=0.9\linewidth]{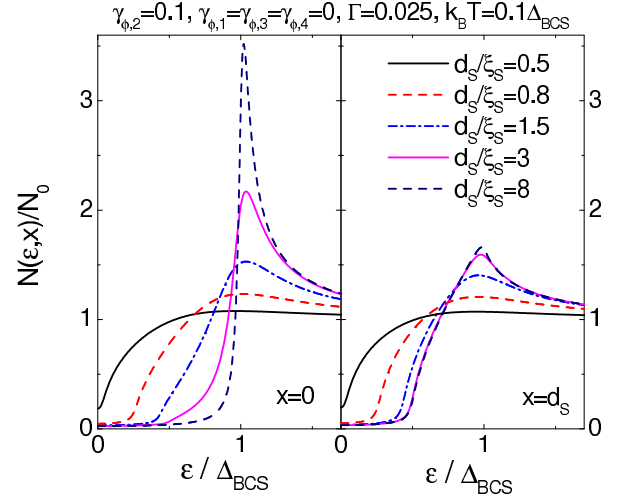}\caption{Density of states in
the $S$ layer contacted to a $FI$, for $x=0$ (left panel) and $x=d_{S}$ (right
panel), and different values of $d_{S}/\xi_{S}$. In all cases, we have used
$\gamma_{\phi,2}=0.1$, $\gamma_{\phi,1}=\gamma_{\phi,3}=\gamma_{\phi,4}=0$,
$\Gamma=0.025$ and $k_{B}T=0.1\Delta_{BCS}$.}%
\label{NUM3}%
\end{figure}First, a weak $G_{\phi,2}$ widens the BCS peak in a way which is
qualitatively different from $\Gamma$ since the curvature of the DOS for
$\varepsilon<\Delta_{BCS}$ has opposite signs in the two cases. Second, even a
very small $\Gamma$ leads to a finite zero-energy DOS, whereas a small
$G_{\phi,2}$ reduces the gap appearing in the DOS but preserves $N(\varepsilon
,x)=0$ for small energies. For larger values of $G_{\phi,2}$, we expect a gap
suppression in the DOS (not shown in Fig. \ref{NUM2}). Note that in Fig.
\ref{NUM2}, for $\gamma_{\phi,2}=0.1$, $\Gamma=0$ and $d_{S}/\xi_{S}=0.5$, the
gap of the DOS would be reduced but still finite (not shown). In these
conditions, using a small $\Gamma$ can trigger the gap suppression as shown by
the black full line in Fig. \ref{NUM3}, right panel. In the $G_{\phi,1}\neq0$
case, the effects of $G_{\phi,2}$ on a thin $S$ remain qualitatively similar,
in particular, the gap reduction in the DOS still occurs, combined with the
$G_{\phi,1}$-induced spin splitting. Figure \ref{NUM1}, left panel compares a
case with $G_{\phi,2}=0$ and a large $\Gamma$ (black dashed line) with a case
with $G_{\phi,2}\neq0$ and a small $\Gamma$ (blue full line). The two cases
can easily be discriminated due to the different curvatures in the DOS.
Importantly, the analogy between a paramagnetic impurity term and the
$G_{\phi,2}$ term is not complete since $G_{\phi,2}$ occurs in the BCIGF
whereas paramagnetic impurities would contribute directly to the Usadel
equation. This discrepancy is revealed by the dependence of the DOS on $d_{S}%
$. Figure \ref{NUM3} presents the DOS at the left and right side of the $S$
layer for different values of $d_{S}$ and $G_{\phi,1}=0$. We obtain a strong
dependence of the $G_{\phi,2}$ DOS-widening on $x$ and $d_{S}$. First, for
$d_{S}/\xi_{S}=0.5$, the DOS at the left and the right sides of $S$ (left and
right panels) are almost identical, with a suppressed gap for the parameters
we consider. When $d_{S}$ increases, the gap reappears in the DOS. For
$d_{S}\gg\xi_{S}$, the DOS at the left side of $S$ tends to the bulk BCS DOS,
with no effect of $G_{\phi,2}$, whereas the DOS at the right side of $S$ still
has a reduced gap. In this limit, one can check that the reduction of the gap
occurs for a slab of $S$ of thickness $\sim\xi_{S}$ near the $S/FI$ interface.
In contrast, paramagnetic impurities would affect the bulk of $S$. Let us now
consider the case $G_{\phi,1}\neq0$ and $d_{S}$ large. In this case, Ref.
\onlinecite{Cottet3} has shown that the $G_{\phi,1}$-induced spin splitting of
the DOS can persist in a slab of $S$ of thickness $\sim\xi_{S}$ near the
$S/FI$ interface. The right panel of Fig. \ref{NUM1} shows an example of DOS
at $x=d_{S}$ for $d_{S}=3\xi_{S}$, in the case $G_{\phi,1}\neq0$, $G_{\phi
,2}=0$ and a large $\Gamma$ (black dashed line), and in the case $G_{\phi
,1}\neq0$, $G_{\phi,2}\neq0$ and a small $\Gamma$ (blue full line). In the
first case, the $G_{\phi,1}$-induced spin splitting of the DOS is not visible
anymore because $H_{eff}$ scales with $1/d_{S}$ and thus becomes too small
compared to the large value of $\Gamma$ used. In the second case, the double
gap splitting is still slightly visible as a cusp in the DOS curve because the
$G_{\phi,2}$ DOS-widening also decreases with $d_{S}$. The effects of the
$G_{\phi,3}$ and $G_{\phi,4}$ terms in the general case will be presented
elsewhere. Before concluding, we note that in circuits enclosing several $FI$
with non-collinear magnetizations and BCS superconductors, it has been found
that the $G_{\phi,1}$ term can induce spin-precession effects which lead to
superconducting correlations between equal spins\cite{Braude}.

\section{Conclusion\label{conclusion}}

To model the behavior of electronic hybrid circuits, a proper description of
the contacts between the different materials is crucial. In this article, we
have derived general boundary conditions relating isotropic Green's functions
at both sides of the interface between two diffusive materials
[Eqs.~(\ref{GBC}), (\ref{GBC3}, and (\ref{der})]. These BCIGF are valid for a
circuit enclosing superconductors, normal metals, and ferromagnets, in a
possibly non-collinear geometry. In general, they require the knowledge of the
full contact scattering matrix, an information usually not available for
realistic interfaces. However, we have shown that in the limit of a specular
tunnel contact with weakly spin-dependent scattering properties, the BCIGF can
be expressed in terms of a few parameters, i.e. the tunnel conductance $G_{T}$
of the contact, a parameter $G_{MR}$ which accounts for the spin-dependence of
the contact scattering probabilities, and four parameters $G_{\phi}^{L(R)}$
and $G_{\chi}^{L(R)}$ which are finite when the contact exhibits a SDIPS
[Eqs.~(\ref{Iprl}) and (\ref{Iprr})]. In the case of a contact with a $FI$
side, we could simplify the BCIGF for a stronger SDIPS [Eq.~\ref{bcSFI}]. We
believe that the various spin-dependent BCIGF derived in this article
represent a solid basis for further developments on superconducting hybrid circuits.

We acknowledge discussions with A. Brataas, B. Dou\c{c}ot, T. Kontos, J.~P.
Morten, and S. Sadjina. This work was financially supported by the DFG through
SFB 513 and SFB 767 and the Landesstiftung Baden-W\"{u}rttemberg (WB). We
acknowledge the hospitality of the Workshop \textquotedblleft Spin and Charge
Flow in Nanostructures\textquotedblright\ at the CAS, Oslo.%

\appendix

\section{Scattering description of a specular and spin-conserving
contact\label{Scattering}}

\subsection{Structure of the electronic scattering matrix\label{Scattering2}}

In this section, we assume that the transverse channel index $n$ and the spin
index $\sigma\in\{\uparrow,\downarrow\}$, corresponding to spin components
along $\vec{Z}$, are conserved when electrons cross the potential barrier
$\bar{V}_{b}$ between the two ballistic zones. In this case, the electronic
scattering matrix $\mathcal{S}^{e}$ is diagonal in the (transverse
channel)$\otimes$spin subspace. The scattering submatrix associated to
electrons with spins $\sigma$ of the n$^{th}$ transverse channel writes%

\begin{equation}
\mathcal{S}_{n\sigma}^{e}=\left[
\begin{array}
[c]{cc}%
r_{L,n\sigma} & t_{R,n\sigma}\\
t_{L,n\sigma} & r_{R,n\sigma}%
\end{array}
\right]
\end{equation}
Here, $r_{L(R),n\sigma}$ denotes the reflection amplitude at side $L(R)$ of
the barrier and $t_{R(L),n\sigma}$ the transmission amplitude from side
$R(L)$\ to side $L(R)$. Flux conservation imposes, for $\sigma\in
\{\uparrow,\downarrow\}$, $\sum\nolimits_{Q\in\{L,R\}}\left(  \arg
(r_{Q,n\sigma})-\arg(t_{Q,n\sigma})\right)  =\pi\lbrack2\pi]$ and
$1-\left\vert r_{Q,n\sigma}\right\vert ^{2}=\left\vert t_{Q,n\sigma
}\right\vert ^{2}=T_{n\sigma}$. In addition, spin-conservation along $\vec{Z}$
allows to map the scattering description of each spin component $\sigma$ onto
a spinless problem. Time reversal symmetry in each of these spinless problems
implies $\arg(t_{L,n\sigma})=\arg(t_{R,n\sigma})$. Therefore, one can use,
without any loss of generality
\[
\mathcal{S}_{n\sigma}^{e}=\left[
\begin{array}
[c]{cc}%
\sqrt{1-T_{n\sigma}}e^{i\varphi_{n\sigma}^{L}} & i\sqrt{T_{n\sigma}%
}e^{i(\varphi_{n\sigma}^{L}+\varphi_{n\sigma}^{R})/2}\\
i\sqrt{T_{n\sigma}}e^{i(\varphi_{n\sigma}^{L}+\varphi_{n\sigma}^{R})/2} &
\sqrt{1-T_{n\sigma}}e^{i\varphi_{n\sigma}^{R}}%
\end{array}
\right]
\]
with $\varphi_{n\sigma}^{L(R)}=\arg(r_{L(R),n\sigma})$. The matrix
$\mathcal{S}_{n\sigma}^{e}$ is entirely determined by $T_{n\sigma}$,
$\varphi_{n\sigma}^{L}$ and $\varphi_{n\sigma}^{R}$. In this article, we use
the parametrization $T_{n\sigma}=T_{n}(1+\sigma P_{n})$ and $\varphi_{n\sigma
}^{L(R)}=\varphi_{n}^{L(R)}+\sigma(d\varphi_{n}^{L(R)}/2)$ [Equations
(\ref{p1}-\ref{p2})].

\subsection{Expression of the transfer matrix with scattering
parameters\label{MS}}

In this section, we assume that the transmission amplitudes $t_{L(R),n\sigma}$
are finite. With the hypotheses made in section \ref{Scattering2}, the
electronic transfer matrix $\mathcal{M}^{e}$ is also diagonal in the
(transverse channel)$\otimes$spin subspace. In the propagation direction
subspace, the submatrix $\mathcal{M}_{n\sigma}^{e}$ has
elements\cite{RefStone}
\begin{equation}
\mathcal{M}_{n\sigma,+,+}^{e}=\left(  t_{L,n\sigma}^{\dagger}\right)  ^{-1}
\label{M--}%
\end{equation}%
\begin{equation}
\mathcal{M}_{n\sigma,+,-}^{e}=r_{R,n\sigma}\left(  t_{R,n\sigma}\right)  ^{-1}
\label{M-+}%
\end{equation}%
\begin{equation}
\mathcal{M}_{n\sigma,-,+}^{e}=-\left(  t_{R,n\sigma}\right)  ^{-1}%
r_{L,n\sigma} \label{M+-}%
\end{equation}%
\begin{equation}
\mathcal{M}_{n\sigma,-,-}^{e}=\left(  t_{R,n\sigma}\right)  ^{-1} \label{M++}%
\end{equation}
We have used above $+/-$ to denote the right/left-going propagation direction.
Using Eqs. (\ref{Mbar}), (\ref{M--}-\ref{M++}) and the parametrization
introduced in section \ref{Scattering2}, one can obtain and expression for the
matrix $\bar{M}$ in terms of $T_{n},P_{n},$ $\varphi_{n}^{L(R)}$ and
$d\varphi_{n}^{L(R)}$. At first order in $P_{n}$ and $d\varphi_{n}^{L(R)}$,
this leads to the expressions (\ref{trail2}-\ref{dX3}).

\section{Calculation of $\check{I}_{L}^{(1)}(\varepsilon)$ for a $S/F$
contact\label{MatrixCurrent}}

In this section, we give details on the calculation of the contribution
$\check{I}_{L}^{(1)}(\varepsilon)$ to the matrix current $\check{I}%
_{L}(\varepsilon)$ to first order in $\delta\bar{X}$. Using Eq. (\ref{PERT2b}%
), one can rewrite Eq. (\ref{matrixcurrentperturbation1}) as
\begin{align}
\check{I}_{L}^{(1)}(\varepsilon)  &  =-4G_{q}\mathrm{Tr}_{n}\left\{  \hat
{T}_{0}\left(  4+\hat{T}_{0}\left[  \left\{  \check{G}_{L},\check{G}%
_{R}\right\}  -2\right]  \right)  ^{-1}\right. \nonumber\\
&  \times\mathrm{Tr}_{s}\left[  \widetilde{W}\right]  \left.  \hat{T}%
_{0}\left(  4+\hat{T}_{0}\left[  \left\{  \check{G}_{L},\check{G}_{R}\right\}
-2\right]  \right)  ^{-1}\right\}  \label{I1}%
\end{align}
The central term
\begin{equation}
\widetilde{W}=\left(  \check{G}_{R}\check{G}_{L}+\hat{Q}_{0}^{-1}\right)
\delta\widetilde{V}\left(  \check{G}_{R}\check{G}_{L}+\hat{Q}_{0}^{-1}\right)
\left(  \mathbf{1}+\hat{\Sigma}_{3}\text{ }\check{G}_{L}\right)
\end{equation}
of this expression can be decomposed as $\widetilde{W}=\sum\nolimits_{j=1}%
^{4}\widetilde{W}_{j}$, with
\begin{align}
\widetilde{W}_{1}  &  =\hat{Q}_{0}\delta\bar{X}\check{G}_{R}\check{G}%
_{L}+\check{G}_{R}\delta\bar{X}^{\dagger}\hat{Q}_{0}\check{G}_{L}\nonumber\\
&  +\check{G}_{L}\check{G}_{R}\delta\bar{X}\hat{Q}_{0}^{-1}+\check{G}_{L}%
\hat{Q}_{0}^{-1}\delta\bar{X}^{\dagger}\check{G}_{R} \label{term1}%
\end{align}%
\begin{align}
\widetilde{W}_{2}  &  =\hat{Q}_{0}\delta\bar{X}\hat{Q}_{0}^{-1}+\check{G}%
_{R}\delta\bar{X}^{\dagger}\check{G}_{R}\nonumber\\
&  +\check{G}_{L}\check{G}_{R}\delta\bar{X}\check{G}_{R}\check{G}_{L}%
+\check{G}_{L}\hat{Q}_{0}^{-1}\delta\bar{X}^{\dagger}\hat{Q}_{0}\check{G}_{L}
\label{term2}%
\end{align}
and $\widetilde{W}_{3(4)}=\widetilde{W}_{1(2)}\hat{\Sigma}_{3}\check{G}_{L}$.
We now develop the trace over the propagation direction\ index $s$ in
Eq.~(\ref{I1}), using expressions (\ref{dX1}), (\ref{Q0}) and (\ref{Q0inv}),
and keeping in mind that $\check{G}_{L}$ and $\check{G}_{R}$ have no structure
in the \underline{$\mathcal{E}$}$\text{ subspace}$. We find \textrm{Tr}%
$_{s}(\delta\bar{X})=$\textrm{Tr}$_{s}(\delta\bar{X}^{\dagger})=$%
\textrm{Tr}$_{s}(\hat{Q}_{0}\delta\bar{X}\hat{Q}_{0}^{-1})=$\textrm{Tr}%
$_{s}(\hat{Q}_{0}^{-1}\delta\bar{X}^{\dagger}$ $\hat{Q}_{0})=0$, so that
\textrm{Tr}$_{s}(\widetilde{W}_{2})=0$. Due to Eqs.~(\ref{Q0}-\ref{Q0inv}), we
find \textrm{Tr}$_{s}(\hat{Q}_{0}\hat{\Sigma}_{3})=$\textrm{Tr}$_{s}(\hat
{Q}_{0}^{-1}\hat{\Sigma}_{3})=0$. Hence, $\delta\bar{X}_{+\text{ }+}$ and the
diagonal elements of $\hat{Q}_{0}$ and $\hat{Q}_{0}^{-1}$ do not contribute to
\textrm{Tr}$_{s}(\widetilde{W}_{1})$. In contrast, the development of
\textrm{Tr}$_{s}(\widetilde{W}_{3(4)})$ involves both $\delta\bar{X}_{+\text{
}+}$ and $\delta\bar{X}_{+\text{ }-}$. We finally obtain
\begin{equation}
\mathrm{Tr}_{s}\left[  \widetilde{W}_{1}\right]  =\left[  \left\{  A,\check
{G}_{R}\right\}  ,\check{G}_{L}\right]
\end{equation}%
\begin{equation}
\mathrm{Tr}_{s}\left[  \widetilde{W}_{3}\right]  =\left[  \left[  C,\check
{G}_{R}\right]  \check{G}_{L},\check{G}_{L}\right]
\end{equation}%
\begin{equation}
\mathrm{Tr}_{s}\left[  \widetilde{W}_{4}\right]  =2\left[  B-\check{G}%
_{R}\left[  \delta\bar{X}_{++},\check{G}_{R}\right]  ,\check{G}_{L}\right]
\end{equation}
with
\begin{equation}
A\left[  F\right]  =\hat{Q}_{0,+-}\delta\bar{X}_{+-}^{\ast}+[-]\hat{Q}%
_{0,-+}\delta\bar{X}_{+-}%
\end{equation}%
\begin{equation}
C=2\hat{Q}_{0,++}\delta\bar{X}_{++}+F
\end{equation}%
\begin{equation}
B=\left(  \hat{Q}_{0,++}^{2}+\hat{Q}_{0,+-}\hat{Q}_{0,-+}\right)  \delta
\bar{X}_{++}+\hat{Q}_{0,++}F-\delta\bar{X}_{++}%
\end{equation}
Expressing $\hat{Q}_{0}$ and $\delta\bar{X}$ in terms of the scattering
parameters $T_{n}$, $P_{n}$, $\varphi_{n}^{L(R)}$, and $d\varphi_{n}^{L(R)}$
[see Eqs.~(\ref{dX1}-\ref{dX3}) and (\ref{Q0})], and developing the trace on
transverse channels in Eq.~(\ref{I1}), we obtain the expression
(\ref{I_1_final}) for $\check{I}_{L}^{(1)}(\varepsilon)$.

\section{General boundary conditions in the normal-state
limit\label{NormalState}}

When there are no superconducting correlations in the circuit, the isotropic
Green's functions $\check{G}_{L(R)}$ write, in the Keldysh space:
\begin{equation}
\check{G}_{L(R)}=\left[
\begin{array}
[c]{cc}%
\check{\tau}_{3} & \check{K}_{L(R)}\\
0 & -\check{\tau}_{3}%
\end{array}
\right]
\end{equation}
In this limit, the elements $\widetilde{D}_{L}^{-1}$ and $\widetilde{D}%
_{R}^{-1}$ appearing in the general BCIGF (\ref{GBC},\ref{GBC3}) take a simple
form. For instance, one finds, in the Keldysh space,
\[
\widetilde{D}_{L}^{-1}=\left[
\begin{array}
[c]{cc}%
\bar{N}_{L} & -\check{\tau}_{3}\bar{N}_{L}\left(  \bar{M}^{\dag}\check{K}%
_{R}\bar{M}-\check{K}_{L}\bar{M}^{\dag}\bar{M}\right)  \bar{N}_{L}\\
0 & \bar{N}_{L}%
\end{array}
\right]
\]
with $\bar{N}_{L}=\left(  1+\bar{M}^{\dag}\bar{M}\right)  ^{-1}$. A similar
expression can be obtained for $\widetilde{D}_{R}^{-1}$ by replacing $\bar{M}$
by $\bar{M}^{-1}$ and $\check{K}_{L[R]}$ by $\check{K}_{R[L]}$. For comparison
with sections \ref{WeakFerro} and \ref{SFI}, we specialize to the case of a
specular contact conserving spins along the interface magnetization.
Equations~(\ref{GBC},\ref{GBC3}) give, for the Keldysh electronic component of
the matrix currents,%
\begin{equation}
\check{I}_{L}^{K,e}(\varepsilon)=2G_{q}\mathrm{Tr}_{n}\left[  -t_{R}\check
{K}_{R}^{e}t_{R}^{\dag}+\check{K}_{L}^{e}-r_{L}\check{K}_{L}^{e}r_{L}^{\dag
}\right]  \label{bn1}%
\end{equation}
and%
\begin{equation}
\check{I}_{R}^{K,e}(\varepsilon)=2G_{q}\mathrm{Tr}_{n}\left[  t_{L}\check
{K}_{L}^{e}t_{L}^{\dag}-\check{K}_{R}^{e}+r_{R}\check{K}_{R}^{e}r_{R}^{\dag
}\right]  \label{bn2}%
\end{equation}
Assuming that the contact is magnetized along $\vec{Z}$, we obtain%
\begin{align}
\check{I}_{L(R)}^{K,e}(\varepsilon)/2  &  =\left(  [G_{T}/2]+G_{MR\text{ }%
}\right)  \check{u}_{\uparrow}\left[  \check{K}_{L}^{e,\uparrow,\uparrow
}-\check{K}_{R}^{e,\uparrow,\uparrow}\right]  \check{u}_{\uparrow}\nonumber\\
&  +\left(  [G_{T}/2]-G_{MR\text{ }}\right)  \check{u}_{\downarrow}\left[
\check{K}_{L}^{e,\downarrow,\downarrow}-\check{K}_{R}^{e,\downarrow
,\downarrow}\right]  \check{u}_{\downarrow}\nonumber\\
&  \mp G_{mix}^{t}\check{u}_{\uparrow}\check{K}_{R(L)}^{e,\uparrow,\downarrow
}\check{u}_{\downarrow}\mp\left(  G_{mix}^{t}\right)  ^{\ast}\check
{u}_{\downarrow}\check{K}_{R(L)}^{e,\downarrow,\uparrow}\check{u}_{\uparrow
}\nonumber\\
&  \pm G_{mix}^{L(R),r}\check{u}_{\uparrow}\check{K}_{L(R)}^{e,\uparrow
,\downarrow}\check{u}_{\downarrow}\pm\left(  G_{mix}^{L(R),r}\right)  ^{\ast
}\check{u}_{\downarrow}\check{K}_{L(R)}^{e,\downarrow,\uparrow}\check
{u}_{\uparrow} \label{Inormal}%
\end{align}
with $\check{u}_{\uparrow(\downarrow)}=1\pm(\check{\sigma}_{Z}/2)$,
\[
G_{mix}^{t}=G_{q}\sum\nolimits_{n}t_{L,n\downarrow}^{\ast}t_{L,n\uparrow}%
\]
and
\[
G_{mix}^{L(R),r}=G_{q}\sum\nolimits_{n}(1-r_{L(R),n\downarrow}^{\ast
}r_{L(R),n\uparrow})
\]
We have checked that in the normal state limit, Eq.~(\ref{IfiGal}) leads to
Eq. (\ref{bn1}) with $t_{R}=0$. Equations (\ref{bn1}), (\ref{bn2}) and
(\ref{Inormal}) are in agreement with the normal-state BCIGF presented e.g. in
Eq.~(2) of Ref.~\onlinecite{Braatas}, up to a prefactor which corresponds to
our conventions\cite{Pref}. Importantly, the derivation of these equations
requires no particular assumptions on the values of $t_{L(R),n\sigma}$ and
$r_{L(R),n\sigma}$. In the normal state-limit, a strong spin relaxation is
often assumed in $F$, so that the $G_{mix}^{t}$ term is disregarded (see e.g.
Eq.~(5) of Ref.~\onlinecite{Braatas}). When the circuit includes
superconducting elements, the expressions of $\widetilde{D}_{L}^{-1}$ and
$\widetilde{D}_{R}^{-1}$ involve e.g. factors $\left(  1+\check{G}_{L}%
^{a(r)}\bar{M}^{\dag}\check{G}_{R}^{a(r)}\bar{M}\right)  ^{-1}$ instead of
$\bar{N}_{L}$. This is why the superconducting BCIGF are difficult to simplify
in the general case.

\section{Equilibrium boundary conditions in the case of superconducting
correlations between opposite spins only\label{GorkovToUsadel}}

This appendix presents the boundary conditions obeyed by the retarded part of
the isotropic Green's functions, in a case where there are superconducting
correlations between opposite spins only. This situation occurs e.g. when all
the ferromagnetic elements of the circuit are magnetized in collinear
directions. For simplicity, we assume that no phase gradient is present in the
system. The conventions chosen in section \ref{general} give, inside conductor
$Q$,
\begin{equation}
\check{G}_{Q}^{r}=\left[
\begin{array}
[c]{cccc}%
\cos(\Lambda_{\uparrow}^{Q}) & 0 & 0 & \sin(\Lambda_{\uparrow}^{Q})\\
0 & \cos(\Lambda_{\downarrow}^{Q}) & \sin(\Lambda_{\downarrow}^{Q}) & 0\\
0 & \sin(\Lambda_{\downarrow}^{Q}) & -\cos(\Lambda_{\downarrow}^{Q}) & 0\\
\sin(\Lambda_{\uparrow}^{Q}) & 0 & 0 & -\cos(\Lambda_{\uparrow}^{Q})
\end{array}
\right]  \label{ang}%
\end{equation}
with $\Lambda_{\uparrow}^{Q}=\Lambda_{\downarrow}^{Q}$ in the spin-degenerate
case. For a metallic contact, using Eqs.~(\ref{der}), (\ref{Iprl}) and
(\ref{Iprr}), one obtains
\begin{align}
-\frac{A}{\rho_{L}}\frac{\partial\Lambda_{\sigma}^{L}}{\partial z}  &
=G_{T}\sin(\Lambda_{\sigma}^{L}-\Lambda_{\sigma}^{R})+iG_{\phi}^{L}\sigma
\sin(\Lambda_{\sigma}^{L})\label{bc1}\\
&  +2i\sin(\Lambda_{\sigma}^{R})\sigma\left(  G_{\chi}^{R}-G_{\chi}^{L}%
\cos(\Lambda_{\sigma}^{L}-\Lambda_{\sigma}^{R})\right) \nonumber
\end{align}
and%
\begin{align}
-\frac{A}{\rho_{R}}\frac{\partial\Lambda_{\sigma}^{R}}{\partial z}  &
=G_{T}\sin(\Lambda_{\sigma}^{L}-\Lambda_{\sigma}^{R})-iG_{\phi}^{R}\sigma
\sin(\Lambda_{\sigma}^{R})\label{bc2}\\
&  -2i\sigma\sin(\Lambda_{\sigma}^{L})\left(  G_{\chi}^{L}-G_{\chi}^{R}%
\cos(\Lambda_{\sigma}^{R}-\Lambda_{\sigma}^{L})\right) \nonumber
\end{align}
\begin{figure}[ptb]
\includegraphics[width=0.7\linewidth]{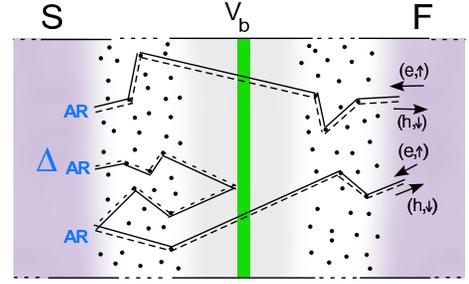}\caption{Scheme
representing two particular types of Andreev reflection processes which can
occur on a $S/F$ interface modelled like in figure \ref{Dessin}. The
ballistic, isotropization and diffusive zones of $S$ and $F$ are represented
by grey, dotted and purple areas respectively. Full (dashed) lines represent
trajectories of electrons (holes) from the $\uparrow$ ($\downarrow$) spin
band. The superconducting gap $\Delta$ is taken into account in the diffusive
part of $S$ only, so that one can consider that Andreev reflections (AR) occur
at the interface between the diffusive and isotropization zones of S. The
upper part of the scheme represents and electron incident from the F side,
which is transmitted by the barrier $V_{b}$ as an electron, Andreev-reflected
on the diffusive part of $S$ as a hole, and transmitted again by $V_{b}$ as a
hole, before joining the diffusive part of $F$ again. The probability
associated to this process is proportional to $T_{n}^{2}(1-P_{n}^{2})$. The
lower part of the scheme represents a more complicated trajectory which also
involves two reflections on $V_{b}$. The joint probability of these
reflections is $(1-T_{n})^{2}-T_{n}^{2}P_{n}^{2}$.}%
\label{AndreevReflexions.eps}%
\end{figure}Interestingly, the $G_{MR}$ term vanishes from Eqs. (\ref{bc1}%
,\ref{bc2}), so that the tunnel rate polarization $P_{n}$ does not contribute
to the equilibrium BCIGF (we have checked that this property remains true when
phase gradients occur in the system). This result may seem surprising since
Andreev reflections, which modify the equilibrium density of states in a
superconducting hybrid system, are suppressed when $P_{n}$ is
strong\cite{DeJong}. However, one should keep in mind that an Andreev
reflection process on the $L/R$ interface involves together the transmission
[or reflection] of an electron \textit{and} a hole from opposite spin bands
through the $\bar{V}_{b}$ barrier (see Figure \ref{AndreevReflexions.eps}).
These two processes have a joint probability which involves $P_{n}^{2}$. In
contrast, single quasiparticle processes, whose probabilities involve $P_{n}$
at first order, do not matter at equilibrium. We conclude that $P_{n}$
vanishes from the equilibrium BCIGF (\ref{bc1},\ref{bc2}) because we have
derived these Eqs. at first order in $P_{n}$. Note that even in this limit,
$P_{n}$ does not vanish from the boundary conditions obeyed by the Keldysh
part of the isotropic Green's functions (see e.g. Eq.~\ref{Inormal}).

For completeness, we mention that in the case of a contact with a $FI$ side,
using Eqs.~(\ref{der}) and (\ref{bcSFI}), one obtains%
\begin{align}
-\frac{A}{\rho_{L}}\frac{\partial\Lambda_{\sigma}^{L}}{\partial z}  &
=iG_{\phi,1}\sigma\sin(\Lambda_{\sigma}^{L})+G_{\phi,2}^{L}\sin(2\Lambda
_{\sigma}^{L})\nonumber\\
&  +iG_{\phi,3}^{L}\sigma\sin(3\Lambda_{\sigma}^{L})+G_{\phi,4}^{L}%
\sin(4\Lambda_{\sigma}^{L})
\end{align}

\section{Usadel Equations\label{UsadelAppendix}}

For completeness, we mention that the Usadel equations corresponding to
Eqs.~(\ref{gorkov},\ref{gorkov2}) write, inside conductor $Q$ (see e.g.
Ref.~\onlinecite {RefWolfgang1})%
\begin{equation}
\hbar D_{Q}\frac{\partial}{\partial z}\left(  \check{G}\frac{\partial
}{\partial z}\check{G}\right)  =\left[  -i\varepsilon\check{\tau}_{3}%
+\check{\Delta}(z)+iE_{ex}(z)\check{\sigma}_{Z},\check{G}\right]
\label{Usadel1}%
\end{equation}
The gap matrix $\check{\Delta}$ has a structure in\ the Nambu-spin subspace
only, i.e., with our conventions,%
\[
\check{\Delta}(z)=\left[
\begin{array}
[c]{cccc}
&  &  & \Delta(z)\\
&  & \Delta(z) & \\
& \Delta(z)^{\ast} &  & \\
\Delta(z)^{\ast} &  &  &
\end{array}
\right]
\]
Using the angular parametrization of section \ref{GorkovToUsadel} and
$\Delta(z)\mathbf{\in}\mathbb{R}$, Eq.(\ref{Usadel1}) leads to:
\begin{equation}
\frac{\hbar D_{Q}}{2}\ \frac{\partial^{2}\Lambda_{\sigma}}{\partial z^{2}%
}=\left(  -i\varepsilon+i\sigma E_{ex}(z)\right)  \sin(\Lambda_{\sigma
})-\Delta(z)\cos(\Lambda_{\sigma}) \label{Usadel}%
\end{equation}


\begin{thebibliography}{99}
\bibitem {def}In this paper, the commutators and anticommutators have a
special meaning when they are applied to the $\mathbf{\Psi}$ and
$\mathbf{\Psi}^{\dagger}$ operators [see Eqs. (3) and (4)], i.e.%
\[
\left\{  \mathbf{\Psi}(1),\mathbf{\Psi}^{\dagger}(2)\right\}  _{\mathcal{T}%
}=\mathbf{\Psi}(1).\mathbf{\Psi}^{\dagger}(2)+\mathcal{T}\left[
\mathcal{T}\left[  \mathbf{\Psi}^{\dagger}(2)\right]  .\mathcal{T}(\left[
\mathbf{\Psi}(1)\right]  \right]
\]%
\[
\left[  \mathbf{\Psi}(1),\mathbf{\Psi}^{\dagger}(2)\right]  _{\mathcal{T}%
}=\mathbf{\Psi}(1).\mathbf{\Psi}^{\dagger}(2)-\mathcal{T}\left[
\mathcal{T}\left[  \mathbf{\Psi}^{\dagger}(2)\right]  .\mathcal{T}\left[
\mathbf{\Psi}(1)\right]  \right]
\]
with $\mathcal{T}$ the transposition operator in the Nambu$\otimes$Spin space.
\bibitem {DefSign}We use the convention \textrm{sign}$(0)=0$ so that the
expression of $\mathbb{G}(\vec{r},\vec{r}^{\,\prime},\varepsilon)$ obtained by
combining Eqs. (\ref{Zaitsev}) and (\ref{match}) is continuous at $\vec
{r}=\vec{r}^{\,\prime}$, in agreement with the definitions (\ref{defG1}%
-\ref{phi}).
\bibitem {scale}Equations (\ref{Gspace}-\ref{Iso4}) confirm that the size
$d_{L(R)}$ of the isotropization zones just needs to be of the order of a few
$\ell_{\text{e}}^{L(R)}$, due to the exponential factors $exp\left[
-z/2\overline{\overline{\upsilon}}_{Q}\tau_{Q}\right]  $. In other terms, only
a few collisions are necessary to isotropize the Green's functions.
\bibitem {QCaverage}The quasiclassical isotropic average of $\mathbb{G}%
(\vec{r},\vec{r}^{\,\prime},\varepsilon)$ can be calculated as $\check
{G}_{qci}(\vec{R},\varepsilon)=i\mathbb{G}(\vec{r}=\vec{R},\vec{r}^{\,\prime
}=\vec{R},\varepsilon)/\pi\nu_{0}$ [see Eq.~(\ref{qci})]. From
Eq.~(\ref{limit}), for $z\rightarrow\mp d_{L(R)}$, one has $\widetilde
{\mathcal{G}}_{diff}(z,z,\varepsilon)=-i\pi\check{G}_{L(R)}$. Inserting this
expression into Eq.~(\ref{Zaitsev}) leads to $\check{G}_{qci}(\vec
{R},\varepsilon)=\check{G}_{L(R)}\mathcal{C}/\nu_{0}$ with $\mathcal{C=}%
\sum_{n}\left\vert \chi_{n}(\vec{\rho})\right\vert ^{2}/\pi\hbar\upsilon_{n}$.
We note that if $t_{L(R),n\sigma}=1$ for any $n$ and $\sigma$, the
normal-state electronic density of states at energy $\epsilon$ for a given
spin direction writes\cite{BlanterButtiker} $N_{\sigma}(\vec{R},\varepsilon
)=\mathcal{C}$, but also $N_{\sigma}(\vec{R},\varepsilon)=\nu_{0}$. The
resulting identity $\mathcal{C}=\nu_{0}$ allows to check, at the beginning of
the diffusive zone $Q$, $\check{G}_{qci}(\vec{R},\varepsilon)=\check{G}_{Q}$.
\bibitem {YipComment}The resemblance between our Eq.
(\ref{Yulicurrent_comutator}) and Eq. (49) of Ref. \onlinecite {Yip} is only
formal because the $\widehat{d}$ operator discussed by Yip for ballistic
systems depends on the ballistic incidence angle. Furthermore, the
angle-average of Eq. (49) of Ref. \onlinecite{Yip} does not yield our
Eq.(\ref{Yulicurrent_comutator}), at least not straightforwardly.
\bibitem {Erratum}In the definitions of $G_{\phi}^{L(R)}$ and $G_{\chi}%
^{L(R)}$ presented in Refs.~\onlinecite {Cottet2,Cottet1} and Fig.~6 of
Ref.~\onlinecite {Cottet2}, the indices $L(R)$ should be changed to $R(L)$.
Then, the definitions of $G_{\phi}^{L(R)}$ and $G_{\chi}^{L(R)}$ given in
these Refs. are correct at zeroth order in $T_{n}$. This does not affect the
results presented in these Refs.
\bibitem {Pref}To check our prefactors, one can consider a $L(R)$ contact in
the normal-state and collinear regime (along $\vec{Z}$), with $L$ and $R$ in
thermal equilibrium at potentials $\mu_{L}$ and $\mu_{R}$, respectively. In
this case, one has $\check{K}_{L(R)}^{e,\sigma,\sigma}=2(1-2f_{L(R)}%
^{e}(\varepsilon))$ and $\check{K}_{L(R)}^{e,\sigma,-\sigma}=0$ for $\sigma
\in\{\uparrow,\downarrow\}$, with $f_{L(R)}^{e}(\varepsilon)$ the equilibrium
distribution function for electrons at side $L(R)$ of the contact (see e.g.
Ref. \onlinecite {wn01}). Combining Eq. (\ref{Inormal}) and the expression
$\left\langle I_{Q}\right\rangle =\int_{-\infty}^{\infty}d\varepsilon
\mathrm{Tr}_{\sigma}\{\check{I}_{Q}^{K,e}(\varepsilon)\}/8\mathrm{e}$ of the
charge current, which derives from Eq. (\ref{current}), one can recover
exactly the Landauer-B\"{u}ttiker expression\cite{BlanterButtiker}
$\left\langle I_{L(R)}\right\rangle =(G_{T}/\mathrm{e})\int_{-\infty}^{\infty
}d\varepsilon\lbrack f_{R}^{e}(\varepsilon)-f_{L}^{e}(\varepsilon)]$ for the
charge current.
\end{thebibliography}
\end{document}